\documentclass[aps,prb,twocolumn,showpacs]{revtex4}
\usepackage[dvips]{graphics}
\newcommand{\Rpi}{\ensuremath{{\cal R}_{\pi}}}
\newcommand{\Hphi}{\ensuremath{\tilde{H}_\phi}}
\begin{document}


\title{ Degeneracy of the ground-state of antiferromagnetic spin-1/2
Hamiltonians.}

\author{G. Misguich}
\email{gmisguich@cea.fr}
\affiliation{
	Service de Physique Th{\'e}orique, Orme des Merisiers\\
	CEA-Saclay F-91191 Gif-sur-Yvette Cedex.}

\author{C. Lhuillier,
        M. Mambrini and P. Sindzingre}
\affiliation{Laboratoire de Physique Th{\'e}orique   des Liquides (UMR  7600  of CNRS)\\
	Universit{\'e} P.  et M. Curie, case 121, 4 Place Jussieu, F-75252 Paris Cedex.}
\date{\today}

\bibliographystyle{prsty}


\begin{abstract}
In   the   first  part   of  this   paper,   the   extension  of   the
Lieb-Schultz-Mattis  theorem to  dimensions  {\em larger  than one} is
discussed.   A    counter  example  to   the  original  formulation of
Lieb-Schultz-Mattis   and Affleck  is   exhibited and  a more  precise
statement  is formulated.   The   degeneracy  of the  ground-state  in
symmetry breaking phases with long-range order is analyzed. The second
and  third parts of   the paper concern resonating valence-bond  (RVB)
spin   liquids.  In  these  phases  the  relationship between  various
authors  approaches:      Laughlin-Oshikawa, Sutherland,   Rokhsar and
Kivelson, Read   and Chakraborty and  the  Lieb-Schultz-Mattis-Affleck
proposal   is studied.   The   deep  physical   relation between   the
degeneracy  property and  the absence of   stiffness  is explained and
illustrated numerically.  A   new conjecture is formed  concerning the
absolute absence of sensitivity  of  the spin liquid ground-states  to
any twist of the boundary  conditions  (thermodynamic limit).  In  the
third part of  the paper the relations between  the quantum numbers of
the degenerate multiplets  of   the spin liquid phases    are obtained
exactly. Their relationship with   a topological property of the  wave
functions   of the  low lying  levels   of this spin   liquid phase is
emphasized.   In spite  of the   degeneracy  of  the ground-state,  we
explain why these phases cannot exhibit spontaneous symmetry breaking.
\end{abstract}
\maketitle
\begin{section}{Introduction}

The   question       of  the degeneracy      of   the  ground-state of
spin-$\frac{1}{2}$ Hamiltonians has a  long history.  One part of this
history goes  back to the beginning of  the sixties when Lieb, Schultz
and Mattis proved that a spin-$\frac{1}{2}$ antiferromagnetic periodic
chain  of   length  $L$  has   a   low   energy excitation   of  order
$\mathcal{O}(1/L)$~\cite{lsm61}.     This  theorem    (called  in  the
following  LSMA) was   then extended  by  Lieb  and  Affleck  to  {\em
half-integer} spin but fails for  integer ones~\cite{al86}.  It states
that in  one  dimension  (1D),  $SU(2)$ invariant   Hamiltonians  with
half-integer spins  in the unit cell,  either have gapless excitations
or  degenerate  ground-states in  the   thermodynamic limit.  The next
important contributions to this long debated question  came at the end
of  the eighties when various authors~\cite{s88,rk88,rc89,h88a} showed
that the ground-state and first excitations  of short-range Resonating
Valence-Bond (RVB) spin  liquids with one spin-$\frac{1}{2}$ per  unit
cell were 4-fold degenerate on a two  dimensional (2D) torus (periodic
boundary conditions).  The connection between the two statements in 1D
was promptly established~\cite{b89a}.  In fact Lieb Schultz and Mattis
had suggested in  their original  paper that  their 1D theorem   could
probably be extended   to higher  space  dimensions:  to support  this
conjecture, they    developed an heuristic    argument,  that has been
refined later on by Affleck~\cite{a88}.

In  this paper  we   examine  the  various  arguments that  have  been
formulated for a 2D spin  systems at the light   of a large number  of
numerical results.

In  section~\ref{section:2DLSMA},   we   study   the situations   with
long-range order  (either   magnetic N{\'e}el  long-ranged   order  or
Valence  Bond Crystals) and  we show that these  systems indeed have a
degenerate  ground-state   in the  thermodynamic   limit. We study two
examples in which  one  aspect of the  LSMA  conjecture appears to  be
incorrect in 2D.

In section~\ref{section:Oshikawa},  we  go through  the  analysis of a
recent  argument by    Oshikawa~\cite{o00}.  We present   a  numerical
analysis   illustrating the   relative   part  of exact   results  and
conjecture in Oshikawa's argument and we show numerically that for the
MSE spin liquid Oshikawa's assumption is verified and the ground-state
is 4-fold degenerate.

In  section~\ref{section:srRVB}   we consider  the   general  case  of
short-range RVB spin liquids (RVBSL).   In such systems the degeneracy
is intimately  related to the existence  of topological sectors in the
Hilbert space of short-range dimer  coverings.  In such a framework we
determine the relations between the  quantum numbers of the degenerate
multiplets both for even$\times$odd  samples and  for even$\times$even
ones.  These results are   illustrated on the quantum  hard-core dimer
(QHCD) model on  the triangular  lattice.  We  finally explain why  in
spin liquids spontaneous  symmetry breaking is  impossible in spite of
the degeneracy of the ground-state.

In the  last section (section~\ref{section:conclusion}),  we summarize
the most  salient results  of the present  work  and  conclude.   Some
technical points are presented in two appendices.

\end{section}

\begin{section}{The LSMA conjecture for 2D systems.} 
\label{section:2DLSMA}

\begin{subsection}{The LSMA theorem for chains and stripes of vanishing
aspect ratio.}
\label{1DLSMA}

The   LSMA theorem applies  to  spectra  of quantum  antiferromagnetic
chains with periodic boundary conditions, and half-integer spin in the
unit cell.  The proof of the theorem is based on the construction of a
low lying excitation, which  may be pictured as   a slow twist of  the
exact ground-state.  Precisely   the excitation is determined   by the
action on the exact ground-state $\left|\psi_0 \right>$ of the unitary
operator $U$ defined as:

\begin{equation}
         U(\phi)= \exp(i\frac{\phi}{L_x}
                {\sum_{j=0}^{L_y -1} \sum_{n=0}^{L_x-1} n  S^z_{n,j}}),
\label{unit_2d}
\end{equation}
where $S^z_{n,j}$ represents the  z component of  the spin operator at
site $(n,j)$, the chain  being of length  $L_x$ and width  $L_y$. The
low energy variational excitation reads:
\begin{equation}
        \left|\theta^{\rm  LSMA}_{2\pi}\right>=U(2\pi)\left|\psi_0\right>.
        \label{LSMA}
\end{equation}
We will design as a  column of the sample  all the spins swept by  the
unit cell subjected to the $L_y$ translations along  $y$.  Let us call
${\cal  T}_x$  the operator for    one-step   translation in the   $x$
direction: the Hamiltonian is supposed to be translationally invariant
$ \left[{\cal H},   {\cal T}_x\right]=0$. It is  easy  to show that  $
U(2\pi)$  anti-commutes with  ${\cal T}_x$  as  soon as the number  of
spins  in  a  column is  an  odd integer~\cite{lsm61,a88}.    In these
conditions, if   the ground-state   wave-vector is ${\bf    k}_0$, the
variational excitation  $\left|\theta^{\rm LSMA}_{2\pi}\right>$  has a
wave-vector   ${\bf  k}_0+(\pi,0)$:  it  is  thus  orthogonal  to  the
ground-state.

The energy of this variational state is:
\begin{equation}
        \left< \theta ^{LSMA} _{2\pi}\right|
        {\cal H}_0\left| \theta ^{LSMA} _{2\pi}\right>
        = E_0  + \alpha L_x L_y  
        \left[1 - cos( \frac{2\pi}{L_x}) \right],
\label{var_energy}
\end{equation}
where $ E_0 =\left<\psi_0\right| {\cal H}_0\left|\psi_0\right>$ is the
ground-state   energy   and  $\alpha$ a   finite   quantity  of  order
$\mathcal{O}(1)$ measuring the   average value in the  ground-state of
spin-spin correlation  functions.   The LSMA   theorem concerns  short
range  Hamiltonians: for the  simple nearest  neighbor Heisenberg case
$\alpha$ is   equal  to  $-2/3$ times  the   energy  per bond   in the
ground-state.

Eq.~(\ref{var_energy}) expresses  the   LSMA theorems: if  the  system
under   consideration  is  a chain,  $\left|    \theta ^{LSMA} _{2\pi}
\right>$ is orthogonal to the ground-state and collapses onto it as ${\cal
O}(1/L_x)$ when the chain length $L_x$  goes to infinity. The property
remains true for a stripe with an odd number of rows and an odd number
of  spins $\frac{1}{2}$ in the  unit cell as soon  as the aspect ratio
$L_y/L_x$ goes to  zero when the size  of the sample goes to infinity:
the energy of the excited state  is then ${\cal O}(L_y/L_x)$ above the
ground-state.  On the other hand, if  the thermodynamic limit is taken
with  a non vanishing aspect ratio,  the  gap between the ground-state
and the   variational  excitation does  not close  in    2D and higher
dimensions.

Faced to   this  alternative, Lieb,  Schultz  and  Mattis~\cite{lsm61}
expressed  the conjecture that  the   degeneracy observed  in the  odd
stripes with vanishing aspect ratio  could be a true physical property
of the 2D samples with aspect ratio equal to 1.  They wrote: ``Because
the excitation  energy of exact  low-lying states should not depend on
the shape of the entire lattice,  there should be  no energy gap for a
lattice of    $N  \times  N$  sites   either''.   Elaborating  on this
consideration,  Affleck~\cite{a88}  expressed essentially    the  same
conjecture.

In the following subsections, we  want to precise these statements  at
the  light of  a numerical  analysis  of the exact spectra  of various
generic  2D quantum antiferromagnets.   To discuss the validity of the
above-mentioned  conjecture for dimensions larger or   equal to 2, we
extract from it different proposals with increasing specificity:
\begin{itemize}
        \item proposal  A:  The ground-state of   an antiferromagnetic
        system with an  odd number of  spins $\frac{1}{2}$ in the unit
        cell is degenerate in the thermodynamic limit.

        \item proposal B: If  ${\bf  k}_0$ is  the wave-vector of  the
        ground-state there are at  least 2 additional eigenstates with
        wave vectors  ${\bf k}_0+(\pi,0)$ and ${\bf k}_0+(0,\pi)$ that
        collapse onto the ground-state in the thermodynamic limit.

        \item proposal ${\rm  B}^{'}$: In the thermodynamic limit  the
        unit cell  of the ground-state is enlarged  by a factor $2^d$,
        where d is the  dimension of the  lattice. This statement is a
        bit  more precise  that  statement B:   in 2D  it  implies the
        degeneracy  of  the  ground-state   with eigenstates  of  wave
        vectors:  ${\bf  k}_0   +(\pi,0)$, $(0,\pi)$  and  ${\bf  k}_0
        +(\pi,\pi)$.

        \item proposal C: A good  variational estimate of the  excited
        states  of proposal  B  can  be obtained  by  a  twist of  the
        ground-state (to be specified in subsection~\ref{subse:propC}).

\end{itemize}

\end{subsection}

\begin{subsection}{Proposal A and spectra of different generic
2D quantum antiferromagnets. }

At  $T=0$ the   ground-state of an   antiferromagnetic  system can  be
N{\'e}el  ordered   (the spin-spin   correlations exhibit   long-range
order),  a Valence-Bond-Crystal   (VBC) (the dimer-dimer  correlations
show   long-range  order)   or   an  RVBSL   state with no  long-range
order~\cite{lm01}.

In any of these  cases  and for  various reasons the  ground-state  is
degenerate in the thermodynamic limit.

In   the first  two-cases the  symmetry  breaking  of the  macroscopic
ground-state  implies   a degeneracy   of    the ground-state in   the
thermodynamic limit.

{\em i)} In   the case of  N{\'e}el long-range  order, there are  both
degenerate states (which form the true thermodynamic ground-state) and
gapless excitations.   The gapless excitations,  the antiferromagnetic
magnons, are  the Goldstone mode  associated to  the broken continuous
SU(2)  symmetry.  Whereas  it is  well known  that the softest magnons
scale as ${\cal  O}  (1/L)$, it is  sometimes  taken for granted  that
these collective ``first excitations'' are the first excited levels of
the multi-particle spectra.   This  is indeed false: the  ``T=0 N{\'e}el
ground-state''  (or  ``vacuum  of excitations'')   is  itself a linear
superposition  of  $\sim  N^{\alpha}$   eigenstates  of   the  $SU(2)$
invariant  Hamiltonian  which     collapse to the     ground-state  as
$\mathcal{O}(1/N)$ ($\alpha$   is the number  of sub-lattices   of the
N{\'e}el state, $L$  is  the linear  size of the  sample and  $N$  the
number of sites of the sample): this is the tower of states invoked in
P.~W.~Anderson 1952 famous paper~\cite{a52,f89,nz89,bllp94}.

{\em ii)} In the case  of a discrete broken  symmetry, as for  example
the  space   symmetry  breaking associated    to long-range  dimer  or
plaquette order (VBC)~\cite{zu96,cs00,sow01,cls00,cms01,fsl01,fmsl01},
or in the case of $T$-symmetry breaking associated to long-range order
in  chirality,  the vacuum  of excitations is  also  degenerate in the
thermodynamic limit.    The spectrum of a  finite-size  sample shows a
quasi-degeneracy of the ground-state.  The  number of quasi degenerate
states is    finite and a  function   of  the symmetry   of  the order
parameter,  it is  independent of  the sample size.    The collapse of
these quasi-degenerate levels on   the ground-state is supposed to  be
exponential with the size of the lattice.

{\em iii)} In the  last case of an RVB  state with no long-range order
we have found two generic situations:

$\alpha)$ A spin liquid with  a gap to all  excitations with up to now
two examples: a phase of the multiple-spin exchange Hamiltonian on the
triangular lattice~\cite{mlbw99},  and    a phase  of  the  frustrated
$J_1-J_2$ model on the honeycomb  lattice~\cite{fsl01} (we shall  call
this phase a type I spin liquid).  Finite  size studies of these cases
point to  a degenerate ground-state  on  the triangular  lattice and a
unique one on the honeycomb lattice.
As  explained in  section~\ref{section:srRVB}, this  difference comes
from the  fact that the triangular lattice  has one spin $\frac{1}{2}$
per unit cell and the honeycomb lattice has 2.

$\beta)$  A spin liquid  with a spin  gap  filled with a continuum  of
singlets on the kagom{\'e} lattice~\cite{lblps97}, and on the triangular
lattice~\cite{lmsl00}. In the last two cases the continuum of singlets
appears to be adjacent to the ground-state (type II spin liquid).

In  any of  these   cases Proposal A   appears  to be   true,  and the
restriction to problems  with half-odd integer spins  in the unit cell
necessary (as can be  seen on the example of  the $J_1-J_2$ model  on
the honeycomb lattice~\cite{fsl01}).

\end{subsection}

\begin{subsection}{Proposal B and ${\rm B}^{'}$: a thorough study of one
 counter-example}

Proposal ${\rm B}^{'}$ is valid when the macroscopic ground-state unit
cell is multiplied by a factor $2^d$.  This is for example the case of
the 4-spin $S=0$ plaquette order of  the $J_1-J_2$ model on the square
lattice for $J_2/J_1\sim 0.5$ \footnote{ The existence  of a VBC phase
for such a coupling is rather well documented, but the exact nature of
the order is still under debate~\cite{zu96,kosw99a,cs00,cbps01}.}.
Proposal  B' is not  valid in the ground-state  of the $J_2/J_1$ model
for $J_2/J_1>>0.5$.  In that case the ground-state is a two-sublattice
columnar antiferromagnet which   spontaneously breaks the  translation
and $\pi/2$ rotation symmetries.  It is easy to check that proposal B'
is not satisfied  in that case since  the spatial symmetry breaking in
that phase is incompatible with momentum ${\bf k}=(\pi,\pi)$. The
four-fold ground-state degeneracy is obtained by a combination of {\em
two} states  with zero momentum, one  with ${\bf  k} =(0,\pi)$ and one
with $(\pi,0)$.

We will now show that even the weaker proposal  B can be false in true
2D  spin systems and  that, contrary to  LSMA assumption,  a 2D sample
with vanishing aspect ratio can have a mathematical spectrum different
from that of  the true 2D  samples.  To illustrate this  assumption we
have  done a  thorough numerical study   of the exact   spectra of the
Heisenberg model  on  the triangular lattice   for  various shapes and
sizes of the samples.

\begin{figure}
        \begin{center}
        \resizebox{8cm}{!}{
        \includegraphics*{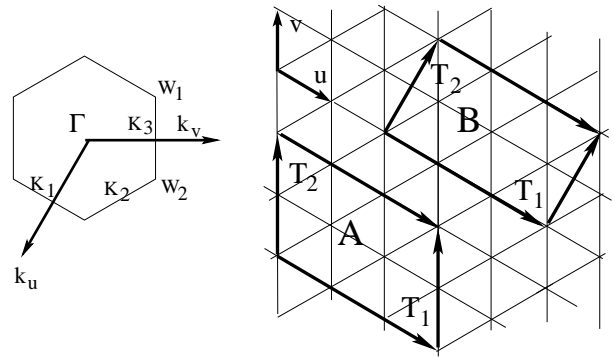}}
        \end{center}
        \caption[99] {   Brillouin  zone of the    triangular lattice:
        wave-vectors ${\bf   K_1}=(\pi,0)$ frequently invoked  in  the
        text are shown  in   the Brillouin  zone.  When  the  $C_{6v}$
        symmetry of   the   sample is  preserved,   eigenstates   with
        wave-vectors    $\bf     K_1$, $\bf   K_2$,   $\bf    K_3$ are
        degenerate. The  two kinds of  samples used  in subsection 2.3
        are designed as A, and B.
        }\label{brillouin}
 \end{figure}

The Heisenberg Hamiltonian with nearest neighbor interactions reads:
\begin{equation}
{\cal H}_0=2 \sum_{<i,j>} {\bf S}_i\cdot{\bf S}_{j}.
\end{equation}   

The Bravais lattice is  defined by the  two unit vectors ${\bf u}$ and
${\bf v}$   . The reciprocal  lattice is  hexagonal:  the  lattice and
Brillouin zone  are displayed  in  Fig.~\ref{brillouin}. The  sample is
defined by two vectors:
\begin{equation}
{\bf T}_{1(2)}  = l_{1(2)}{\bf u}  + m_{1(2)}{\bf v}.
\end{equation}
Periodic boundary conditions read:
\begin{equation}
{\bf S}({\bf R}_i+{\bf T}_j)=
        {\bf S}({\bf R}_i)
\label{pbc}
\end{equation}   
For a given sample the wave-vectors compatible with the
boundary conditions depend on ${\bf T}_1, {\bf T}_2$.

Previous studies on samples with aspect ratio equal to 1~\cite{bllp94}
have established that this model has  a three sublattice N{\'e}el order.
The Anderson tower of states  collapsing to the absolute  ground-state
as $1/N$ encompass eigenstates with wave-vectors: $ {\bf k =0}$ and \\
$ {\bf  k}  = \pm (2\pi/3,2\pi/3)$.   The low-lying  magnons have wave
vectors  in the neighborhood  of the points  $\Gamma$  and $W_i$ of the
Brillouin  zone of the system (see  Fig.~\ref{brillouin}).  Due to the
overall consistency of the picture, we suspected that eigenstates with
wave-vectors   $(0,\pi)$    or  $(\pi,0)$  (points    ${\bf   K_i}$ of
Fig.~\ref{brillouin}) appear only in components of energetic magnons.

In this work we report  a thorough numerical study of the behavior of
the first  ${\bf k}  =  {\bf k}_0  +(\pi,0)$ exact eigenstate  of this
model as a function of the aspect ratio and size of the samples (${\bf
k}_0$ is  always the wave vector of  the absolute ground-state for the
sample under consideration).

We  have studied samples  with  various  values of $(L_x,L_y)$  (even,
odd),  (even,   even)  accommodating  eigenstates   with   wave-vector
$(\pi,0)$. For a  given aspect ratio  two kinds  of samples have  been
studied:
\begin{itemize}
        \item The simplest one has its sides parallel to the primitive
        vectors of  the lattice: ${\bf  T}_1  =L_x{\bf u},  {\bf T}_2=
        L_y{\bf v}$. If $L_x$ or $L_{y} = 1,2 \bmod 3\  $, PBC on such
        samples frustrate the natural N{\'e}el order of the lattice: the
        ground-state is   usually not a   ${\bf   k=0}$ state and  the
        spectrum is not typical of N{\'e}el order. We nevertheless study
        these samples which  are  perfectly  reasonable  samples  with
        respect     to    the    LSMA   conjecture     (A  sample   of
        Fig.~\ref{brillouin}).
        
        \item If one dimension only is not multiple of  3, one may use
        a different shape, with  the same aspect  ratio, where PBC  do
        not frustrate    N{\'e}el    order     (see  B      sample    of
        Fig.~\ref{brillouin}   and   double    starred  samples     in
        Table~\ref{table-1}).  On  these non frustrating samples (NFS)
        the absolute ground-state   is a ${\bf   k=0}$  state and  the
        spectrum  has the    typical  structure of   a N{\'e}el  ordered
        system.  On such samples  one can  also build the  variational
        state described   by Eq.~(\ref{LSMA}).  It has   a wave-vector
        $(\pi,0)$, and  its  gap   to the absolute    ground-state  is
        $3\alpha L_y/L_x$  (to be compared with Eq.~(\ref{var_energy})
        for an A sample).
\end{itemize}

\begin{table}
\begin{center}
\begin{tabular}{|c|c c|c|c|c|c|c| }
\hline
$N$&$L_x$&$L_y$&$\Delta_{s}$&$\Delta_{0}$&$\Delta_{1}$&$\delta$ & \\
\hline
12*  &  $\sqrt12$  & $\sqrt12$   &  1.726  &  2.471  &  2.836  & 2.47 & no\\
\hline
18*  &  6  &  3  & 1.131  & 1.483  & 2.882 & 2.96 & no \\
\hline
 24  &  8  &  3  & 0.942  & 0.474  & 1.486 & 1.26 & yes \\
   24  &  6  & 4  &  1.193  &  1.424  & 1.697 & 2.13  & no\\
   24**  &  8  & 3  &  0.926  & 1.034  &  2.073 & 2.75 & no \\
   24**  &  6  & 4  &  0.922  & 1.061  &  2.175 & 1.59  & no\\
\hline
   30  &  10 & 3  &  0.799  & 0.344  &  1.183 & 1.15 & yes \\
   30  &  6  & 5  &  0.825  & 0.762  &  1.652 & 0.91 & yes \\
   30**  &  10  & 3  & 0.782  & 0.732  & 1.618 & 2.43 & yes \\
   30**  &  6  & 5  &  0.776  & 1.067  & 2.003 & 1.28 & no \\
\hline
   36* & 12 & 3 & 0.692 & 0.548 & 1.324 & 2.19 & yes \\  36* & 6 & 6 &
   0.688 & 1.359 & 1.785 & 1.36 & no \\
\hline
\end{tabular}
\end{center} 
        \caption[99]{ Gaps   of  various   excited states   (from  the
        absolute ground-state):  $\Delta_{s}$, spin gap; $\Delta_{0}$,
        gap between the absolute ground-state  (with wave vector ${\bf
        k_0}$)  and the first   excited  state with wave  vector ${\bf
        k_0}+ (\pi, 0)$ in the $S=0$ sector; $\Delta_{1}$, gap between
        the absolute ground-state and the first state with wave vector
        ${\bf k}_0 +  (\pi, 0)$ in the  $S=1$ sector; $\delta$, is the
        ratio  of   $\Delta_{0}$ to the  aspect   ratio  of the sample
        $L_y/L_x$.  In the last column we indicate if $\left|{\bf k}_0
        + (\pi,0), S=0\right>$  is  the  first excited state   of  the
        spectrum. The lines with one or two *  indicate A or B samples
        with PBC compatible  with 3-sublattice N{\'e}el long-range order
        (See Fig.~\ref{brillouin} for the double-starred samples).  }
\label{table-1}
\end{table}

 \begin{figure}
        \begin{center}
        \resizebox{8cm}{!}{
        \includegraphics*{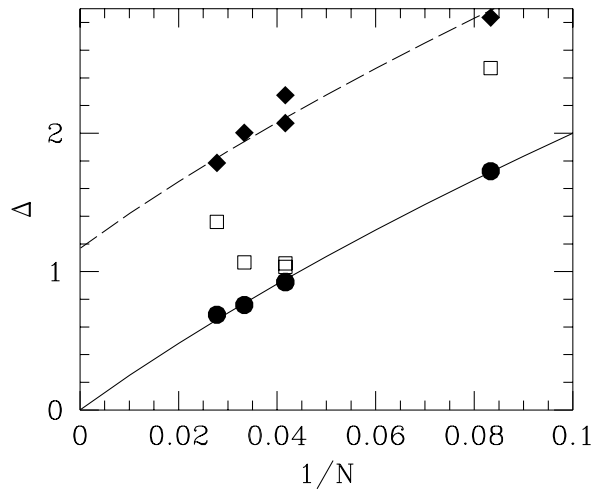}}
        \end{center}

        \caption[99] {Finite-size  scaling of various gaps  to excited
        states.  Black circles:    $\Delta_s$ gap to  the  first $S=1$
        state (it has wave-vector $W_{1(2)}$ in  the Brillouin zone of
        Fig.~\ref{brillouin}).  Open squares:   $\Delta_0$ gap to the
        first $\left|(\pi,0),  S=0\right>$ state (points $K_i$  in the
        Brillouin zone).  Black  diamonds: $\Delta_1$ gap to the first
        $\left|(\pi,0),   S=1\right>$     state.    }\label{proposalB}
        \end{figure}

This    study    leads  to       the   following  observations    (see
Table~\ref{table-1}):
\begin{itemize}
        \item For  a  given size  the spin  gap $\Delta_s$ has  a weak
        dependence with   the sample shape (it  is  especially weak on
        NFS).  It decreases  with the  sample size  as  expected for a
        N{\'e}el ordered system (see Fig.~\ref{proposalB}):
        \begin{equation}
                \Delta_s= \frac{1}{ \chi N} (1 - \beta\frac{c}{\rho \sqrt{N}}) 
                + {\cal O} (\frac{1}{N^2})
                \label{eq-spin-gap}
\end{equation}
        where $\chi$  is the  homogeneous spin susceptibility,  $c$ the
        spin-wave velocity, $\rho$ the spin stiffness and $\beta$ is a
        number  of order one. The  first triplet excited states of NFS
        have wave-vectors
        ${\bf k} = \pm (2\pi/3,-2\pi/3)$ and ${\bf k} = {\bf 0}$.

        \item The gap $\Delta_0$ to the first excited state with total
        spin   $S=0$   and wave-vector  ${\bf  k}_0  +(\pi,0)$ behaves
        differently.  (This state  is the first   exact state with the
        same wave-vector as $\left|\theta^{LSMA}_{2\pi}\right>$).

        \item $\Delta_0$ is  small for small  aspect  ratio and for  a
        given size is  very sensitive to the aspect  ratio and to  the
        boundary conditions. For a  small aspect  ratio this state  is
        indeed   the first excited  state of  the exact spectrum (last
        column of Table~\ref{table-1}).

        \item For a given size and  aspect ratio, $\Delta_0$ increases
        when going from frustrating  periodic boundary  conditions, to
        boundary  conditions  compatible  with   the three  sublattice
        N{\'e}el  order   (except for the   $6 \times   4$ sample, where
        $\Delta_0$ is always large).

        \item  For a given size, $\Delta_0$  increases with the sample
        aspect  ratio:  whatever the size  of  the  sample this gap is
        always of the same order of magnitude as the aspect ratio (6th
        column of Table~\ref{table-1})

        \item Let us consider now  the two samples  with $6 \times  5$
        and  $6 \times 6$  sites which do  not  frustrate N{\'e}el order
        (double      starred         and      starred   lines       of
        Table~\ref{table-1}). When going  from  the N=30   spectrum to
        $N=36$ all  gaps   decrease, which seems natural,   except the
        $\Delta_0$ gap which increases  noticeably.  This is a  second
        indication of  the strong dependence of  the position  of this
        specific level on the sample's aspect ratio.

        \item We  finally looked at the size  and shape effects on the
        gap $\Delta_1$  to the first  state with wave-vector $(\pi,0)$
        and  total  spin $S=1$:  $\left|(\pi,0), S=1\right>$   (Let us
        recall that  $\left|\theta^{\rm  LSMA}_{2\pi}\right>$ has not a
        fixed   total spin). For  any   size   and shape $\Delta_1   >
        \Delta_0$.   $\Delta_1$  data  corresponding to  NFS  with the
        largest aspect ratios obey the same finite size scaling law as
        the $\Delta_s$ spin gap (Fig.\ref{proposalB}): this is exactly
        what is expected for a $(\pi,0)$ magnon excitation (the magnon
        excitation could be written  as a twist of a symmetry-breaking
        ground-state, which itself is a  coherent superposition of the
        low     lying   levels   of      the    Anderson    tower   of
        states). Extrapolation  to the   thermodynamic limit gives   a
        energy of $1.17$ (the linear spin wave  theory gives the value
        2, but all  the ED data point to  a softening of the spin-wave
        spectrum by  quantum  fluctuations\footnote{This softening due
        to quantum fluctuations always noticeable on the stiffness and
        spin-wave  velocity, seems even  larger on  the maxima of  the
        dispersion curves, as  it  has also be  noticed on  the square
        lattice~\cite{r01}.}).

        \item It  is difficult to assign a  value to the thermodynamic
        limit  of   the gap  $\Delta_0$   to the first $\left|(\pi,0),
        S=0\right>$ state  for NFS with  the largest aspect ratio (see
        Fig.\ref{proposalB}) (two samples only, 12  and 36 have aspect
        ratio equal to 1): however examination  of all the data pleads
        in favor of a  non zero gap. If  the semi-classical picture is
        valid,  the  state $\left|(\pi,0), S=0\right>$ belongs  to the
        tower of states of the  one-magnon  excitation of wave  vector
        $(\pi,0)$, and   in the  thermodynamic  limit \begin{equation}
        lim_{N\to        \infty}\left(\Delta_0\right)        =
        lim_{N\to \infty}\left(\Delta_1\right). \end{equation}
        The present data are too scarce to sustain this last assertion
        but they do not go against it.
\end{itemize}

This study suggests that:

{\it i)} The thermodynamic limit: \\ $L_x,L_y \to \infty$ with
$L_y/L_x \to 0$  is special, it could  depend on the number of
rows in the sample (see sample $6 \times 4$) and is different from the
limit: $L_x,L_y \to \infty$ with $L_y/L_x =\mathcal{O}(1)$.

{\it ii)} The  first state $\left|(\pi,0), S=0\right>$ which collapses
to  the  ground-state  in the  quasi one-dimensional  system  is  not a
``natural'' excitation  of the 2D  system: the natural excitations are
the magnons and  the magnon with   the wave-vector $(\pi,0)$  is not a
soft mode as soon as  the period of the symmetry-breaking ground-state
is not even (as in the case above, where the period is 3).

{\it  iii)} The nearest   neighbor Heisenberg model  on the triangular
lattice  appears to  be a serious   counter-example to proposal B  and
${\rm B}^{'}$ for a true 2D system. We suspect  that the same model on
the  square  lattice   would equally be   a  counter-example  to these
proposals.  We  have not done a full  size and  shape analysis of this
case as we have done for the triangular lattice.  But we have verified
that  for samples with  non zero  aspect  ratios the Anderson tower of
states encompasses eigenstates with  wave-vectors ${\bf  k}=(0,0)$ and
${\bf k}=(\pi,\pi)$ as expected, and the ${\bf k}= (\pi, 0)$ and ${\bf
k}= (0,\pi)$ are components of  {\rm energetic} magnons~\footnote{ The
first  excited state of the spectrum  is an $S=1$ ${\bf k}= (\pi,\pi)$
eigenstate: its gap to the ground-state goes to zero linearly with the
size of the  sample;  its value is 0.630  (resp.   0.575) in  the N=32
(resp.   N=36) sample.  The first  eigenstate with a wave-vector ${\bf
k}= (\pi, 0)$ has a gap to the  ground-state of 4.885 (resp. 4.850) in
these two samples: it  is triplet ($S=1$).   The first eigenstate with
$S=0$ and ${\bf  k}= (\pi, 0)$  has a huge  gap to the ground-state of
5.195 (resp. 5.075). }.

\end{subsection}

\begin{subsection}{Proposal C} \label{subse:propC}

Eq.~(\ref{var_energy})   implies that  in a  true  2D  system it is in
general impossible  to  build a low-lying   excitation using  the LSMA
twist operator  $U(2\pi)$.  In Ref.~\cite{a88}, Affleck  discussed the
general  possibility  to  build a   one-dimensional   excitation in  a
two-dimensional medium.   We do not  know how to do  this in a general
case, and  as explained in  the  previous subsection, we  suspect that
this quest is in general doomed to failure.

However, in the specific case of spin liquids, it could be possible to
build such an excitation, inasmuch as the stiffness of a 2D spin liquid
vanishes when the size of the sample goes to infinity.

In an   ordered  system, the  stiffness  $\rho$,   which  measures the
response of  the sample $L_x  \times L_y$  to a   twist $\phi$ of  the
boundary condition in the $x$ direction, can be defined as
\begin{equation}
        E(\phi)= E(0) + \frac{N}{2} \rho
        \left(\frac{\phi}{L_x}\right)^2 + {\cal O}(\phi^4).
        \label{stiffness}
\end{equation}
with $E(\phi)$ the total  energy of the  sample.  Far from the quantum
critical point the  stiffness is usually of the  order of magnitude of
the coupling constant  ($\sim  1/9$ for  the  Heisenberg model on  the
triangular lattice).

Let us now consider the following  one-dimensional perturbation of the
ground-state with PBC: a constraint on  line number one is forced with
the twist operator
\begin{equation}
        u^{1}(2\pi)= \exp(i\frac{2\pi}{L_x}\sum_{n=0}^{L_x-1} n  S^z_{n,1})
        \label{unit_restrict}
\end{equation}
and the  system is  supposed  to  relax around  this default  at fixed
momentum (Eq.~(\ref{unit_restrict}) insures that the variational state
has momentum $(\pi,0)$).

In  an ordered system with  stiffness $\rho$ the variational energy of
such an excitation could be approximated as:
\begin{equation}
        E \sim E_0   +
        \frac {1}{2}\;\beta \rho \frac{L_y}{L_x}  \left(\pi \right)^2, 
        \label{ener_stripe}
\end{equation} 
with $\beta$ a number  of order one.  In  a true 2D ordered system the
correction  to the  ground-state energy  can  never go  to zero  for a
non-zero aspect ratio.  On the contrary, in a spin liquid sample, with
linear dimensions larger than  the  correlation length, we  conjecture
that  the second  term of  the  above estimate  goes to  zero with the
stiffness.

This heuristic argument is in agreement with the behavior of the exact
spectra  described  in  the  next  sections.  The   key  ingredient to
discriminate between  the two cases  is  the presence of  a finite (or
vanishing)   stiffness: a fact     intimately related to the   central
hypothesis of Oshikawa that we will discuss in the following section.
\end{subsection}
\end{section}

\begin{section} {Oshikawa's argument}
\label{section:Oshikawa}

In a  recent paper~\cite{o00} Oshikawa   discussed the consequences of
the presence of a gap in a quantum many-particle system with conserved
particle    number on a   periodic  lattice  in arbitrary  dimensions.
Inspired by Laughlin's  topological  arguments  for the  quantum  Hall
effect, Oshikawa argues that a finite  excitation gap is possible only
when the  particle number  per  unit cell  of  the ground-state  is an
integer.  The above-mentioned formulation translated for a spin system
reads: ``A spin  system   can have  a  gap only   if the spin  in  the
ground-state  unit cell  is   integer'' and Oshikawa  statement  for a
spin-$\frac{1}{2}$ system is  ``a spin-$\frac{1}{2}$ system can have a
gap only if, the translational symmetry is spontaneously broken'': the
ground-state  unit cell is thus  adequately  enlarged to be consistent
with the previous statement.

\begin{figure}
        \begin{center}
        \resizebox{3cm}{!}{\includegraphics{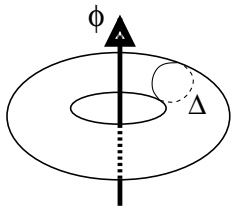}}
        \end{center} \caption{2-torus with one cut $\Delta$.}
        \label{torus} 
\end{figure}

To  support his   statement Oshikawa  developed   an argument  {\`a}  la
Laughlin involving  the adiabatic transform  of the ground-state under
the action   of a fictitious magnetic   flux piercing the  ring of the
sample submitted  to periodic  boundary conditions (Fig.~\ref{torus}).
In  this section  we  illustrate this adiabatic   transform on  a spin
system. This  transform is   here realized in  the  spin   language by
twisting           the           boundary                   conditions
(subsection~\ref{subsec:adiabatic_transform}).       We  then show the
general   mathematical properties of the spectrum   of  {\it any} spin
Hamiltonian     under      such         an    adiabatic      transform
(subsection~3.2 and Appendix A).

Oshikawa's result  is based on a crucial  hypothesis that the gap from
the ground-state manifold to excited state {\em  does not close during
the adiabatic  transform}.  In  subsection~\ref{GapRobustness}      we
illustrate  this assumption  on exact   spectra  of the  multiple-spin
exchange (MSE)  Hamiltonian, in a range  of parameter where this model
exhibits a RVBSL phase in the  thermodynamic limit (see Appendix B for
the precise definition of the model and the  range of parameters where
it exhibits a  spin liquid phase).   We postpone the discussion of the
second issue of Oshikawa's paper, i.e.   the discussion of spontaneous
symmetry breaking to the next section.

\begin{subsection}{Adiabatic transform of a spin-system under the twist
of the boundary conditions.}
\label{subsec:adiabatic_transform}

If  the  spin-$\frac{1}{2}$  degrees  of freedom   are  represented by
hard-core  bosons  (Holstein-Primakoff), coupling   these bosons  to a
fictitious  flux is completely equivalent, in  the spin language, to a
{\em twist} of  the boundary conditions.  Generalized twisted boundary
conditions are defined  by the choice of the  twist axis (here the $z$
axis in the original ${\cal B}_0$ spin frame)  and a set of $2$ angles
$\phi_j$ as:
\begin{equation}
{\bf S}({\bf R}_i+{\bf T}_j)=e^{i\phi_jS^z({\bf R}_i)}
        {\bf S}({\bf R}_i)
        e^{-i\phi_jS_{z}({\bf R}_i)}
\label{twbc}
\end{equation}
(see Fig.~\ref{brillouin} for the definitions of $T_1$ and $T_2$). For
the sake  of simplicity we  will illustrate  the general properties of
this  transform on  the case  of a twist  $\phi$  in the ${\bf T}_{1}$
direction for the simplest $SU(2)$  invariant Hamiltonian: the nearest
neighbor  Heisenberg  Hamiltonian (extension   to a more   complex and
relevant Hamiltonian is just  a question of notations).  With periodic
boundary   conditions the Hamiltonian  in  the ${\cal B}_0$ spin frame
reads:
\begin{equation}
        {\cal H}_0=\sum_{p=0}^{L_y -1}\sum_{n=0}^{L_x-1}
        \left( {\bf S}_{n,p}\cdot{\bf S}_{n+1 [L_x],p}
        +{\bf S}_{n,p}\cdot{\bf S}_{n,p+1[L_y]}\right)
\end{equation}
A twist $\phi$ in the  ${\bf T}_{1}$ direction implies the calculation
of the eigenstates of
\begin{eqnarray}
        {\cal H}_{\phi}&=&\sum_{p=0}^{L_y-1}\left({\bf S}_{L_x-1,p}\cdot
        {\bf S}_{0,p}^{\phi}
        +{\bf S}_{L_x-1,p}\cdot{\bf S}_{L_x-1,p+1[L_y]}\right)  \nonumber \\
        && +
        \sum_{p=0}^{L_y-1}\sum_{n=0}^{L_x-2} \left({\bf S}_{n,p}\cdot{\bf S}_{n+1,p}
        +{\bf S}_{n,p}\cdot{\bf S}_{n,p+1[L_y]}\right) \nonumber \\
        &=&{\cal H}_0
        +\frac{1}{2}\sum_{p=0}^{L_y-1}
        \left( ( e^{i\phi} -1) S_{L_x-1,p}^-S_{0,p}^++{\rm h.c.}\right)
        \label{Hphi}
\end{eqnarray} 
where
\begin{equation}
        {\bf S}_{0,p}^{\phi}= e^{i\phi S^z_{0,p}}{\bf S}_{0,p}e^{-i\phi S^z_{0,p}}.
\end{equation}
The $\phi$ twist  term in Eq.~(\ref{Hphi})  is equivalent to the phase
factor induced by  a magnetic  flux  on  the kinetic  term of  charged
particles in  Laughlin's transform.  Under an  adiabatic  twist of the
boundary conditions the spectrum of ${\cal H}_\phi$ (Eq.~(\ref{Hphi}))
evolves  periodically with  a   period $2\pi$.   But the   eigenstates
evolution could be more complicated.
There  is  no guaranty   that  the  many-body ground-state  of  ${\cal
H}_{\phi =0}$ adiabatically transforms into the ground-state of ${\cal
H}_{\phi = 2\pi}$.   As we will show  below, this is generally not the
case and the true periods of the  eigenstates are in general multiples
of $2\pi$.
\end{subsection}

\begin{subsection}{Twisted boundary conditions and translation invariance}

To  follow adiabatically an  eigenstate  of  a many particle  spectrum
while twisting the boundary conditions could seem difficult because of
level  crossings  during the  twist.    In fact there  is  an  unique,
unambiguous way   to   do    it as   these  eigenstates    belong   to
one-dimensional  representations  of  a translation   symmetry  group.
Diagonalization in such subspaces  leads to non degenerate eigenstates
which never cross.   The translational  invariance  of the  problem is
hidden in  its  representation in the   ${\cal B}_0$ frame  but can be
restored by a gauge transform.

Let  us   define ${\cal B}_{\phi}$,  the  new  frame deduced  from the
original  ${\cal B}_0$ by  the spatially  dependent twist described by
the unitary operator:
\begin{equation}
        U(\phi)= \exp( i\frac{\phi}{L_x}
                {\sum_{p=0}^{L_y-1}\sum_{n=0}^{L_x-1} n  S^z_{n,p}}
        )
        \label{unit}
\end{equation}
In this new frame, the twisted Heisenberg Hamiltonian reads:
\begin{equation}
\tilde{{\cal H}}_\phi=U(\phi) {\cal H}_{\phi} U(\phi)^{-1};
\end{equation}
(in  this   equation and in  the  following,  we put  a  tilde  on the
observables measured in the ${\cal B}_\phi$ frame. We will indicate by
a superscript $\phi$ the eigenstates of $\tilde{{\cal H}}_\phi$). This
unitary   transform  is   chosen  so  that  the     boundary terms  in
Eq.~(\ref{Hphi}) disappear:
\begin{equation}
        \tilde{{\cal H}}_\phi=
        {\cal H}_{0}
        +\frac{1}{2}{\sum_{p=0}^{L_y-1}\sum_{n=0}^{L_x-1}\left((e^{i\phi/L_x}-1)
{S}_{n,p}^-{S}_{n+1 [L_x],p}^++{\rm h.c.}\right)}.
\label{h1h0}
\end{equation}
$\tilde{{\cal H}}_\phi$ is translation invariant: $ \left[\tilde{{\cal
H}}_\phi, {\cal T}_{x(y)} \right]=0$,  where ${\cal T}_{x(y)}$ are the
operators for one-step translations in the $x$ (resp. $y$) directions.
Its spectrum   is  indeed   identical to   the  spectrum   of   ${\cal
H}_\phi$. After  this   gauge    transform,    we can   now     define
one-dimensional irreducible  representations  of the translation group
labeled by their wave-vectors  ${\bf k}$ in  the ${\cal B}_\phi$ frame
and  adiabatically  follow a   ${\bf k}$  eigenstate  of $\tilde{{\cal
H}}_\phi$  in the successive  ${\cal B}_\phi$  frames while increasing
$\phi$.  This is what is done in Figs.~\ref{LROtwspectra} and \ref{twistedspectrum}, where we see
that the true period of  the  many-body ground-state $|{\bf  k}=(0,0),
S^z_{tot}=0>^{\phi}$   and   of  the  first exact   $|{\bf k}=(\pi,0),
S^z_{tot}=0>^{\phi}$ state of the $8 \times 3$ sample (resp. of the
$4 \times 5$ one) is $4\pi$.
\begin{figure}
        \begin{center}
        \resizebox{8cm}{!}{
                \includegraphics{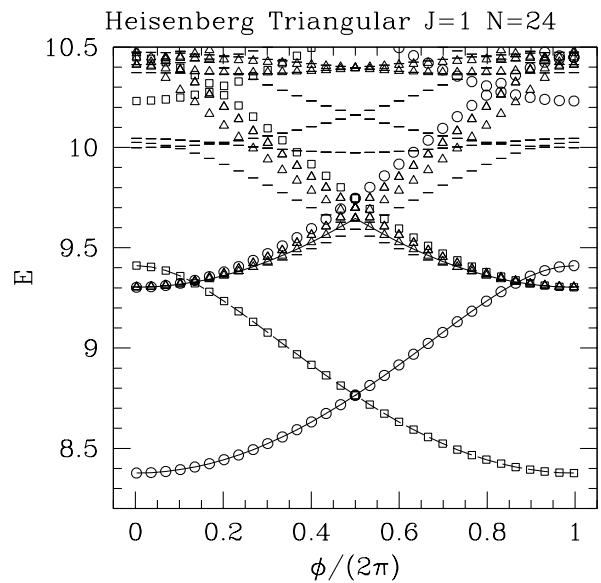}}
        \end{center}

        \caption{Low-lying spectrum  of the $8\times  3$ sample of the
        N{\'e}el ordered Heisenberg model on the triangular lattice as a
        function  of   the twist $\phi$  at  the  $x$  boundary of the
        sample. Circles are for $|S^z_{tot}=0,{\bf k}={\bf 0}>^{\phi}$
        states  and squares stand  for states with $|S^z_{tot}=0, {\bf
        K}_{1}>^{\phi}$ (wave-vector $(\pi,0)$).   Triangles represent
        the   lowest    $S^z_{tot}=1$  state    (whatever    its  wave
        vector). Continuous  lines    are  drawn through   the   first
        $|S^z_{tot}=0,{\bf    k}={\bf 0}>^{\phi}$  state,  the   first
        $|S^z_{tot}=0,    {\bf   K}_{1}>^{\phi}$    and    the  lowest
        $S^z_{tot}=1$ state thus giving  the  spin  gap.  Due to   the
        finite   size, the  ${\bf   k}$ wave   vector  of  this lowest
        $S^z_{tot}=1$ state  changes with the twist angles (explaining
        the  discontinuities in the slope  of the  curve) and with the
        sample size. } \label{LROtwspectra}
\end{figure}
\begin{figure}
        \begin{center}
        \resizebox{8cm}{!}{
                \includegraphics{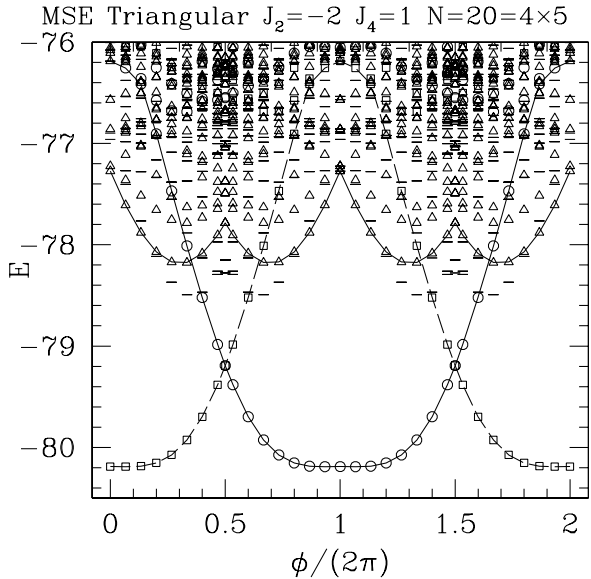}}
        \end{center}

        \caption{Low-lying spectrum of the $4 \times  5$ sample of the
        multiple-spin exchange  model (Eq.~(\ref{MSE})) as a function of
        the  twist $\phi$ at  the  $x$ boundary   of the sample.  Same
        symbols  as in  Fig.~\ref{LROtwspectra}.  We suspect  that for
        this spin liquid, in  the thermodynamic limit the  wave vector
        of  the first $S^z_{tot}=1$ state  is $(\pi/2,0)$.  But in the
        finite small  samples  that are  available we  never have this
        wave  vector (except in  the $4\times6$ sample): this explains
        the  discontinuities in the shape  of the lowest $S^z_{tot}=1$
        state.} \label{twistedspectrum}
\end{figure}

\begin{figure}
        \begin{center}
        \resizebox{8cm}{!}{
                \includegraphics{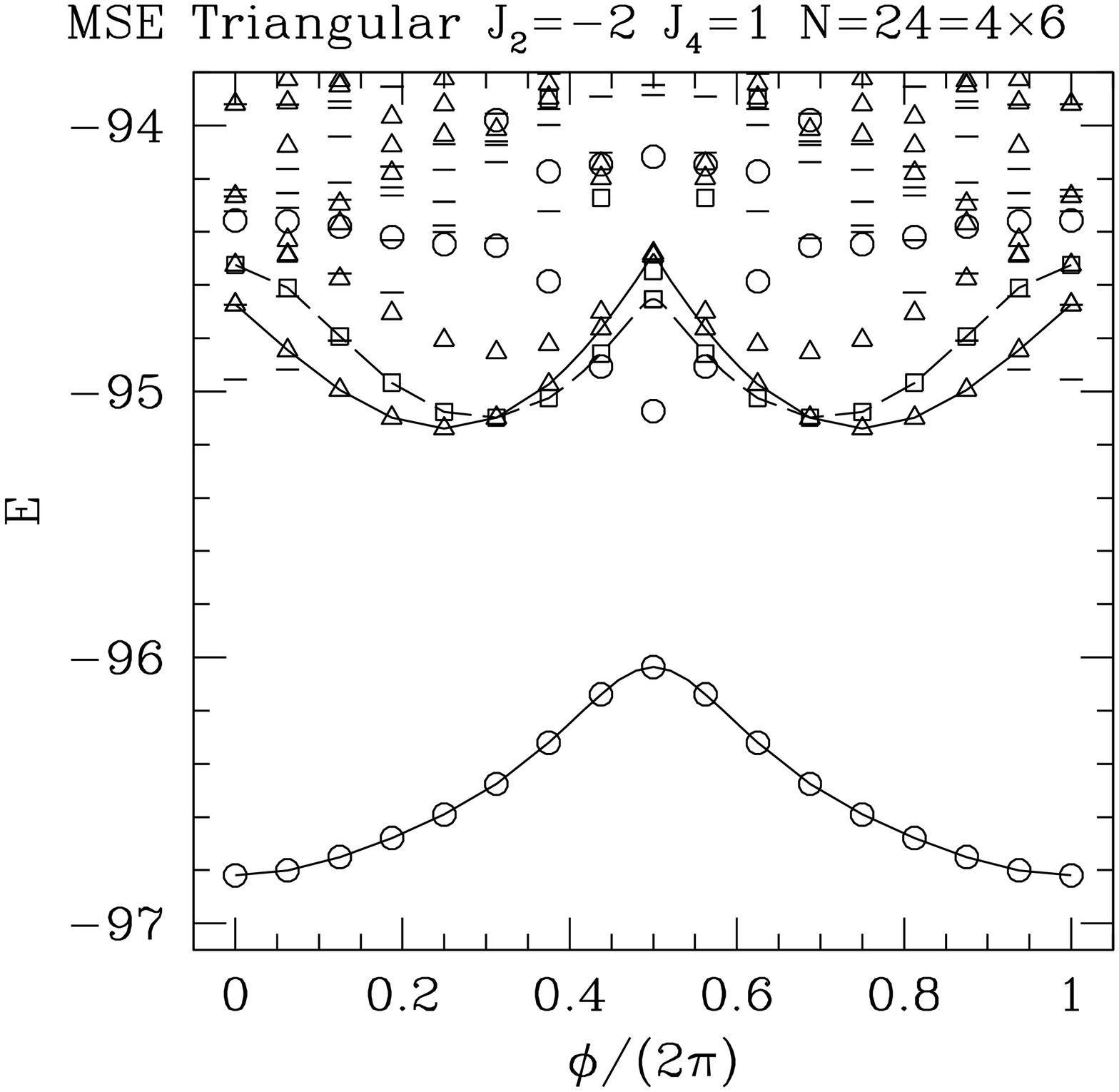}}
        \end{center}

        \caption{Low-lying spectrum of the $4  \times 6$ sample of the
        multiple-spin exchange model  (Eq.~(\ref{MSE}))  as a function
        of  the twist $\phi$ at the  $x$ boundary of  the sample. Same
        symbols as in Fig.\ref{LROtwspectra}.}
\label{scan4x6}
\end{figure}

\begin{figure}
        \begin{center}
        \resizebox{8cm}{!}{
                \includegraphics{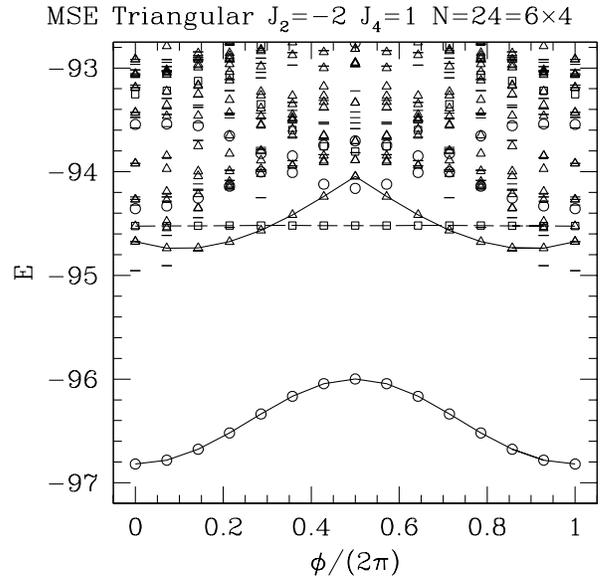}}
        \end{center}

        \caption{Low-lying spectrum  of the $6\times  4$ sample of the
        multiple-spin exchange  model (Eq.~(\ref{MSE}))  as a function
        of the  twist $\phi$ at the $x$  boundary  of the sample. Same
        symbols as in Fig.\ref{LROtwspectra}.  }
\label{scan6x4}
\end{figure}

This property is an exact property of the $S^z_{tot}=0$ eigenstates on
even$\times$odd samples.   A straightforward  algebra (computation  of
${\cal  T}_{x(y)} U(2\pi){\cal T}^{-1}_{x(y)}$),  shows that $U(2\pi)$
maps  a state with  wave-vector ${\bf  k}$  and total spin $S^z_{tot}$
onto a state with the same $z$-component of  the total spin but with a
wave vector ${\bf k'}$ given by~\footnote{This  relation has also been
used to  generalize to $S_{\rm  tot}^z\neq0$ the LSMA argument in order
to discuss magnetization plateaus in one dimension~\cite{oya97}.}:
\begin{eqnarray}
 k_x' &=&  k_x  +
        2\pi\frac{S_{\rm tot}^z}{ L_x}
        + \pi \left[L_y\;{\rm mod}\;2\right] \nonumber \\
 k_y' &=&  k_y.
\label{k1k2}
\end{eqnarray}

From these relations one immediately deduces that a $2\pi$ twist on an
even$\times$odd sample maps the absolute ground-state $|{\bf k}=(0,0),
S^z_{tot}=0>$  of  $  {\cal H}_0$ onto   the  first $|{\bf k}=(\pi,0),
S^z_{tot}=0>$  eigenstate and reversely, {\it  whatever  the nature of
the Hamiltonian and  the intrinsic  properties of its   ground-state}.
Fig.~\ref{LROtwspectra}  is a spectrum of  a  N{\'e}el ordered phase and
Figs.~\ref{twistedspectrum} and ~\ref{twist30} of spin liquid phases.

More generally the $S^z_{tot}=0$  eigenstates have periodicity  $4\pi$
on   even$\times$odd     samples     (see    Figs.~\ref{LROtwspectra},
\ref{twistedspectrum}  and~\ref{twist30})          and      $2\pi$  on
even$\times$even  ones (Figs.~\ref{scan4x6},  \ref{scan6x4})  . Let us
finally remark that all eigenstates of a given problem do not have the
same periodicity:  the periodicity depends both on  the z component of
the  total  spin  of the  eigenstate   under consideration and  on its
wave-vector   (Eqs.~(\ref{k1k2})).   Additional   information  on  the
quantum numbers  associated   with the point-group   symmetries can be
found in appendix A.

\begin{figure}
        \begin{center}                             \resizebox{8cm}{!}{
        \includegraphics{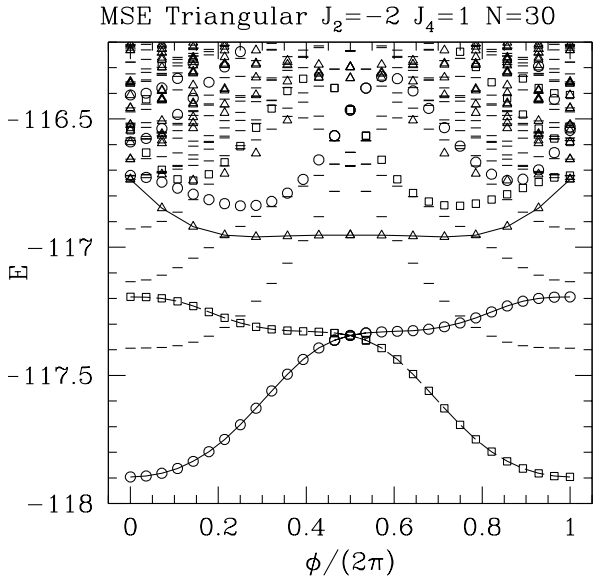}}  \end{center}   \caption{
        Low-lying spectrum  of  the    $6\times  5$  sample  of    the
        multiple-spin exchange model  (Eq.~(\ref{MSE})) as  a function
        of the twist $\phi$ at  the $x$ boundary  of the sample.  Same
        symbols as  in Fig.~\ref{LROtwspectra}.  On this  small sample
        the lowest  $S^z_{tot}=1$  state  is   a  $|S^z=1, {\bf    k}=
        (2\pi/3,0)>$.  We suspect  that in the thermodynamic limit the
        lowest $S^z=1$ eigenstate has  a wave-vector belonging  to the
        star of $(\pi/2,0)$ (these wave-vectors are not allowed in the
        present sample).} \label{twist30}
\end{figure}

\end{subsection}

\begin{subsection}{
Robustness of the  gap in presence  of twisted boundary  conditions in
spin liquids
}\label{GapRobustness}

To  establish the existence  of a ground-state  degeneracy in a gapped
system, Oshikawa makes the  following assumption: the ground-state  of
${\cal H}$ can    only  be transformed in    itself  or into   another
degenerate  ground-state.  In other words,    he assumes that the  gap
 from the ground-state manifold to   excited states does not
close during the  adiabatic evolution.  Thanks to  the  results of the
previous  sub-section,  this  hypothesis directly    implies that   in
even$\times$odd   samples the  $|S^z_{tot}=0,{\bf  k}=(0,0)>$ absolute
ground-state is     degenerate with  the   lowest   $|S^z_{tot}=0,{\bf
k}=(\pi,0)>$ eigenstate.   {\it If    we take  for  granted   that  in  the
thermodynamic   limit      the   spectra    of even$\times$odd     and
even$\times$even samples are  identical}, Oshikawa's assumption implies
Proposal B'
\footnote{
 In fact we will show below that the four-fold degeneracy of the RVBSL
 can  sometimes be  realized  in even$\times$odd  and even$\times$even
 samples  with different irreducible   representations  of the   space
 group. }.

Numerical results for the  RVBSL phase of the  MSE model support these
two hypotheses   (see Appendix B   for the precise description  of the
model and spectral results in Figs.~\ref{twistedspectrum},
\ref{scan4x6}, \ref{scan6x4},
\ref{twist30}  and Table~\ref{table-2}). The  gap  $\Delta_0$ between the absolute
ground-state   $|S^z_{tot}=0,{\bf     k}=(0,0)>$   and    the    first
$|S^z_{tot}=0,{\bf k}=(\pi,0)>$  decreases steadily  with the size  of
the   sample with     no   difference   between   even$\times$odd  and
even$\times$even   samples    (see       Table~\ref{table-2}       and
Fig.~\ref{gaps_size}).
  
\begin{table}
\begin{center}
\begin{tabular}{|c|c|c|c|c|c|}
\hline
N&20&24&28&30&36\\
\hline
$\Delta_{0}$&4.0027&2.2944&1.5683&0.7026&0.2527\\
\hline
$\Delta ^{'}_{0}$&0.9967&0.8195&0.6958&0.5615&\\
\hline
$\Delta _S$&2.9168&2.1452&1.2555&1.1607&0.9847\\
\hline
\end{tabular}
\end{center} 
\caption[99]{Spin liquid spectra (MSE model defined in Appendix B). 
$\Delta_{0}$ measures the gap  between the absolute ground-state  with
wave vector  $(0,0)$,  and the  first  excited state with  wave vector
$(\pi,0)$ in the $S=0$ sector; $\Delta ^{'}_{0}$ measures the increase
in the absolute  ground-state total  energy in a  $\pi$ twist  of  the
boundary conditions; $\Delta _S$ is the spin gap.  }
\label{table-2}
\end{table}

These results pull into light the central conjecture of Oshikawa, {\it
i.e.}    what  Laughlin   calls the  robustness   of   the  gap or the
uncompressibility of  the spin  liquid and  its relationship  with our
conjecture of the  vanishing stiffness of the  2D spin liquids  in the
thermodynamic limit.  Strictly  speaking numerical calculations do not
allow a consistent determination of the ``stiffness'' of a finite size
sample in  a spin liquid phase:  depending on the  shape of the sample
and on the direction of the  applied twist the  response of the system
can vary in a  range going from half  to one hundredth of the coupling
constant (contrary to the case of a N{\'e}el  ordered system where the
stiffness  is   very  well defined   and   totally  independent on the
direction of the twist).  Most of the results obtained  on the 28, 30,
32-sites samples  are   10 to 100   times smaller  than those   of the
N{\'e}el ordered  phase.  But the conjecture  that can be  formed from
the above mentioned results is even stronger than the vanishing of the
stiffness and  amounts to {\it an absolute  absence  of sensitivity of
the thermodynamic ground-state of  a spin liquid  to any twist  of the
boundary conditions} ({\it i.e.} all terms  of the Taylor expansion of
Eq.~(\ref{stiffness}) are  zero in the  thermodynamic  limit, not only
the $\phi^2$ term).

This conjecture  is   based on the following   observations concerning
spectra    of    the   MSE   model    (see    Table~\ref{table-2}  and
Figs~\ref{twistedspectrum}, \ref{scan4x6},
\ref{scan6x4}, \ref{twist30}):
\begin{itemize}
        \item in even$\times$odd  samples  the ground-state energy  of
        the system under a  twist oscillate between  the energy of the
        absolute  ground-state at zero twist   and  that of the  first
        excited state  $|S^z_{tot}=0,{\bf k}=(\pi,0)>$ reached  for  a
        $2\pi$ twist\footnote{In small frustrated samples the absolute
        minimum could  be observed for  a very small, non zero, twist.
        This peculiarity is no more present for the stablest symmetric
        samples of size larger or equal to 28.}.  The amplitude of the
        oscillation measured by  the  energy  gap $\Delta_0$   between
        these two states decreases steadily with the system size.

        \item In even$\times$even samples the amplitude of oscillation
        is reduced:   the  maximum of  energy  of  the ground-state is
        obtained for a $\pi$  twist. The amplitude of this oscillation
        $\Delta'_0$ given  by the  difference in total  energy between
        these   two states  is given  in  Table~\ref{table-2}.  As the
        previous gap, this amplitude decreases with the system size.
        
        \item  The number of samples   is  too small  to ascertain  an
        exponential   decrease to zero  with  system  length, but this
        conjecture remains plausible (Fig.~\ref{gaps_size}).
\end{itemize}

 \begin{figure}
        \begin{center}
        \resizebox{8cm}{!}
        {\includegraphics*{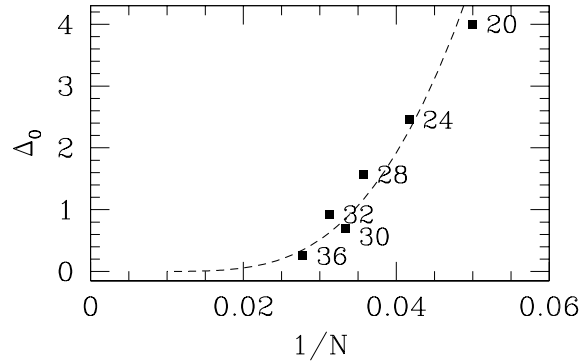}}
        \end{center}
        \caption[99]  { $\Delta_0$ gap of the   first multiplet in the
        singlet sector  versus $\sqrt N$: the   line is an exponential
        fit                               $\propto\exp(-\sqrt{N}/\xi)$
        ($\xi\sim0.6$)}\label{gaps_size}
\end{figure}
On physical grounds,   the absence of any response   to a twist   is a
rather natural property in an RVBSL since there such  a phase offer no
local order parameter this perturbation could couple to.
In section~\ref{section:srRVB} we will come back to this question from
the point of view of the short-range RVB picture of spin liquids.

\end{subsection}

\begin{subsection}{Oshikawa's conjecture and Valence Bond Crystals}

The  numerical work presented in  the  previous subsection concerns an
RVBSL. We  have not investigated the   behavior of a VBC  with twisted
boundary conditions but it is clear that the stiffness should be zero.
In principle non-linear effects could be different in VBC and RVBSL.

\end{subsection}
\end{section}

\begin{section}{The short-range RVB picture of the ground-state
multiplicity of a spin liquid}
\label{section:srRVB}

Because of  the short-range and  antiferromagnetic nature  of magnetic
correlations and    spin-rotation  invariance,  short-range      dimer
configurations offer  a natural  description of low-energy  degrees of
freedom in  spin liquids.  Indeed,  much  insights into their physical
properties come  from the investigation of   dimer models. The quantum
hard    core    dimer     (QHCD)     model    of       Rokhsar     and
Kivelson~\cite{rk88,s88,rc89}  on the  square  lattice is the simplest
dimer model but one has  to consider the triangular lattice  (Moessner
and  Sondhi~\cite{ms01})  to find   a   genuine liquid  with  a finite
correlation  length.  The short-range dimer point  of view also gave a
significant  insights on  another  different kind of  spin liquid: the
Kagome Heisenberg antiferromagnet (see Ref.~\cite{mm01} and references
therein).

First (\ref{subsec:4sectors})  we  recall that  the Hilbert  space  of
short-ranged dimer  coverings on a   torus  is made  of 4  topological
sectors which cannot  be coupled by any local  operator.  We show that
any wave  function described in a  short-range dimer basis is two-fold
degenerate   on  large enough  even$\times$odd    and can  be  two- or
three-fold  degenerate  on even$\times$even  samples with  appropriate
point-group symmetries.   This is illustrated  on spectra of  the QHCD
model on the triangular lattice.  Then (\ref{sec:4fdeg}) we argue that
for a  spin liquid  where all  dimer  correlations decay exponentially
fast with distance this  degeneracy is extended  to 4 for large enough
samples.   This property is checked   on the  QHCD   model and in  MSE
spectra.

After some remarks  on the  four-fold degeneracy  in the  liquid phase
(\ref{remarks})  we  explain  in paragraph~\ref{sec:no_sym_break}  why
RVBSL,   in spite of  the degeneracy   of  their  ground-states cannot
manifest spontaneous symmetry breaking.

\begin{subsection}{Decomposition of the space of short-range dimer
coverings in topological subspaces and exact degeneracies}
\label{subsec:4sectors}

The space of  nearest neighbor dimer coverings of  a torus is a  small
fraction of  the total  $S=0$ sector of  the full  Hilbert space which
increases as   $\alpha^N$  (with   $\alpha  =1.1753,2^{1/3}, 1.3385,\\
1.5351$ respectively on the hexagonal, kagom{\'e}, square and triangular
lattices). These different coverings are not orthogonal, but for large
enough sizes they form a set of linearly independent vectors: an exact
demonstration    exists    for  the     coverings    of    the  square
lattice~\cite{cck89}, numerical studies have  shown that the  property
is  true for the kagom{\'e} lattice~\cite{mm01},   and in the triangular
lattice for any size large enough to insure than any periodic image of
a site is at a distance larger than 4 units (this work).
   
\subsubsection{Definition of the topological sectors}
\label{deftopsec}
Let us  draw a cut $\Delta$ encircling the torus created by periodic
boundary   conditions  (see Fig.~\ref{torus}).   This hyper-surface of
dimension $d-1$ cuts bonds of the lattice but there is no site sitting
on it. The position  of the cut is  arbitrary.  We may decide in order
to follow closely our previous discussion to put it between spin $L-1$
and  spin    $0$ of each   row   of   the lattice.   The   family   of
nearest-neighbor dimer coverings  can be decomposed into two subspaces
${\cal E}^\pm_\Delta$ depending on the    parity $\Pi_\Delta$ of   the
number  of    dimers    crossing the cut    $\Delta$~\footnote{Another
equivalent definition is based on winding numbers of transition graphs
with a reference configuration~\cite{rk88}.}.  By considering a set of
$d$ cuts $\Delta_{i=1,\dots,d}$ encircling  the torus in all  possible
directions one obtains $2^d$ families of dimer covering.

Any movement of dimers  can be represented  as  a set of closed  loops
around  which  dimers are   shifted in a  cyclic way.    A {\em local}
operator  will only generate contractible  loops which will cross each
cut a {\em  even} number of times.   The number of dimers crossing the
cut can therefore only be changed by an  even integer and the parities
$\Pi_{\Delta_i}$ are unchanged.
        
This  property remains true as  long as one  works in a subspace where
the  dimer lengths are  smaller than  the linear  system size, that is
when  the topological sector are  well  defined (if a  dimer length is
half the   linear of the system   one cannot decide  by which  side it
goes).  On the other hand, we checked on the triangular (resp. Kagome)
lattice that these 4 sectors are  the only topological sectors: local
3- (resp 4-) dimer moves can be used to transform a configuration of a
given sector into any other configuration of that sector.
          
We  show   that these subspaces are   orthogonal  in the thermodynamic
limit~\cite{b89a}.  The graph of the scalar product $<c^+|c^->$ of two
dimer configurations belonging to different subspaces ${\cal E}^+$ and
${\cal E}^-$  has at least one long  loop encircling the  torus in the
$L_x$ direction.  When $L_x$ goes to infinity this contribution to the
scalar product is smaller than  $2^{-L_x/2}$.  Consider two normalized
vectors  $\left|\Psi^+\right>$  and  $\left|\Psi^-\right>$ belonging to  two
different sectors:
\begin{equation}
	\left|\Psi^\pm\right>=\sum_{c^\pm\in\mathcal{E}^\pm}\Psi^\pm(c^\pm) \left|c^\pm\right>
\end{equation}
Because of the exponential number  of dimer coverings in each  subspace
it is not obvious that $\left|\Psi^+\right>$  and  $\left|\Psi^-\right>$ are
orthogonal in the thermodynamic limit. It is indeed the case:
\begin{eqnarray}
\left|\left<\Psi^+|\Psi^-\right>\right|&=&\left|\sum_{c^+,c^-} {
{\Psi^+}^*(c^+)}\Psi^-(c^-)<c^+|c^->\right|
\label{ineqA}\\
&\leq&2^{-L/2}
	\sum_{c^+,c^-} \left| { {\Psi^+}^*(c^+)}\Psi^-(c^-) \right| \label{ineqB}\\
&\leq&2^{-L/2}
	\sqrt{
	\sum_{c^+}\left|\Psi^+(c^+)\right|^2  \sum_{c^-}\left|\Psi^-(c^-)\right|^2
	} \\
&\leq&2^{-L/2} \| \Psi^+ \|\cdot \|\Psi^-\|
\end{eqnarray}
where the  norm in the last  inequality refers to  the diagonal scalar
product $<c_{1}||c_{2}>=\delta_{c_1,c_2}$. We now wish to show that these later
bound is  finite. In order to see  that it is  the case one can expand
the usual scalar product $<c_1|c_2>$ in  powers of $x=2^{-1/2}$ in the
spirit of Ref.~\cite{rk88}. The zeroth order term is nothing but the
diagonal scalar
product:
\begin{equation}
	<c_1||c_2>=\delta_{c_1,c_2} + {\cal O}(x^4)
\end{equation}
The convergence of this loop expansion implies that $\|\Psi\|$ is finite if
$|\left<\Psi|\Psi\right>=1$   and   therefore   $\left<\Psi^+|\Psi^-\right>={\cal
O}(2^{-L/2})$. From this we see that not  only scalar products of {\em
dimer  configurations}   in   different  topological    sectors vanish
exponentially but this is also true for any pair of {\em states}.

In  the following, unless explicitly   mentioned, we consider the  2D
case for simplicity but most of the topological arguments about dimer
covering immediately extend to higher dimensions.

\subsubsection{Two-fold degeneracy in even$\times$odd samples}

In the special  case of tori  with an odd number  of rows (and  an odd
number  of spin-$\frac{1}{2}$ per crystallographic  unit cell),  one step
translation along the $x$ axis (called  ${\cal T}_x$ in the following)
maps  ${\cal E}^+$ on  ${\cal  E}^-$ and  reversely. Some  point-group
symmetry can  also do this job. A  $\pi$ rotation about a lattice site
nearby the  cut (called $\Rpi$ in  the following) has the same effect.
If the cut is chosen  parallel to a symmetry  axis  of the sample,  a
reflection   with  respect to   this axis   (called  $\Sigma_y$ in the
following)  will equally  map   ${\cal E}^+$   on  ${\cal  E}^-$   and
reversely.

All  these symmetry operations isolate a  single column $C$ of lattice
sites  between $\Delta$ and  its transform  $\Delta'$.   In that  case
columns have an odd number of  sites and an odd  number of dimers must
connect  some sites  inside  $C$ with   sites  outside $C$.  Therefore
$\Pi_{\Delta}$ differs from   $\Pi_{\Delta'}$ and  the two   subspaces
${\cal E}^+$ and ${\cal E}^-$ are exchanged.

For  a  large enough system  these  two sectors  are 1) orthogonal, 2)
uncoupled by any   local  Hamiltonian and  3) exchanged   by  symmetry
operations   (even$\times$odd).  This is  enough to  insures that they
have the  same  spectrum, irrespectively of the   physics of the model,
provided it can be described in the short-range  dimer space. In fact,
quantum  numbers of these doublets  of degenerate  states are fixed by
symmetry.

We decompose an eigenstate $|\psi_0>$ on the two topological subspaces
defined relatively to the cut $\Delta$:
\begin{equation}
        \left| \psi_0\right> =\left|\psi_0^+\right>
        +\left|\psi_0^-\right>
\end{equation}
where  $\left|\psi_0^{\pm}\right>$   belong respectively   to the sets
${\cal E}_\Delta^{\pm}$.  $\left|  \psi _{0}\right>$, as an eigenstate
of  the Hamiltonian with periodic boundary  conditions,  belongs to an
irreducible representation of  the translation group. In the following
we  will also  assume  an   $\Rpi$  and $\Sigma_y$  invariance of  the
Hamiltonian and that,   for simplicity, $|\psi_0>$  transforms under a
one-dimensional representation under $\Rpi$ and $\Sigma_y$.

\begin{eqnarray}
        {\cal T}_x \left| \psi_{0}\right> =
        e^{i{\bf k}_0\cdot{\bf u}}\left|\psi_{0}\right>\nonumber\\
        \Rpi \left| \psi_{0}\right> = \rho^{\pi}_0 \left| \psi_{0}\right>\nonumber\\
        \Sigma^{y} \left| \psi_{0}\right> = \sigma^y_0 \left| \psi_{0}\right>
\end{eqnarray}
In       the  thermodynamic     limit  $\left|\psi_0^+\right>$     and
$\left|\psi_0^-\right>$ are orthogonal and   ${\cal T}_x$, $\Rpi$  and
$\Sigma^y$ map ${\cal E}^+$ on ${\cal E}^-$ and reversely:
\begin{eqnarray}
        {\cal T}_x \left| \psi_0^\pm\right> =
                e^{i{\bf k}_0\cdot{\bf u}}\left| \psi_0^\mp\right>\nonumber\\
 \Rpi \left| \psi_0^\pm\right> =
                \rho^\pi_0 \left| \psi_{0}^\mp\right>\nonumber\\
 \Sigma^y \left| \psi_0^\pm\right> =
                \sigma^y_0 \left| \psi_{0}^\mp\right>.
        \label{nq}
\end{eqnarray} 
Let us now build the variational state:

\begin{equation}
\left|\psi_{1,\Delta}\right> =
        \left|\psi_{0}^+\right> - \; \left|\psi_{0}^-\right>.
        \label{eq:psi1Delta}
\end{equation}

Eqs.~(\ref{nq}) imply:
\begin{eqnarray}
        {\cal T}_x \left| \psi_{1,\Delta}\right>
                = - e^{i{\bf k}_0\cdot{\bf u}}\left| \psi_{1,\Delta}\right>\nonumber\\
        \Rpi \left| \psi_{1,\Delta}\right>
                = -\rho^{\pi}_0 \left| \psi_{1,\Delta}\right>\nonumber\\
        \Sigma^{y}\left| \psi_{1,\Delta}\right>
                = -\sigma^y_0 \left| \psi_{1,\Delta}\right>
\end{eqnarray}
$\left| \psi_{1,\Delta}\right>$ has thus a wave-vector ${\bf k}_1$, a
rotation quantum number $\rho_1^\pi$ and a reflection quantum number
$\sigma^y_1$ related to the quantum numbers of $\left|\psi_0\right>$
by the relations:
\begin{eqnarray}
        {\bf k}_1 = {\bf k}_0 +(\pi,0)\nonumber\\
        \rho^{\pi}_1= - \rho^{\pi}_0\nonumber\\
        \sigma^y_1 = - \sigma^y_0.
        \label{symmetryqn}
\end{eqnarray}
It  is  thus  a  state  orthogonal to   the   ground-state (even  on a
finite-size  system where the topological  sectors are not rigorously
orthogonal).

Since  any   local  Hamiltonian has  exponentially   vanishing  matrix
elements between different sectors we have
\begin{equation}
	\left<\psi_0^+\right|
	{\cal H}_0
	\left|\psi_0^-\right> \to 0
\end{equation}
and    $|\psi_{1,\Delta}>$  is   thus degenerate    with the  absolute
ground-state,        their      symmetries    being    related      by
relations~(\ref{symmetryqn}).

We have computed numerically the spectrum of  the QHCD model for three
different  triangular  samples:  $N=20=4\times5$, $N=28=4\times7$  and
$N=30=6\times5$   sites.  These  samples   are translation and  $\Rpi$
invariant  (but without symmetry  axis). As  expected, the spectrum is
exactly    two-fold     degenerate.   Pairs    of  states    satisfying
Eq.~(\ref{symmetryqn}) are exactly degenerate.  The two ground-states,
in  particular, have  ${\bf  k}=(0,0)$ $\Rpi=1$ and  ${\bf k}=(\pi,0)$
$\Rpi=-1$.

\subsubsection{Degeneracies in even$\times$even samples}
\label{sss:exact_deg_e_e}

As remarked  by  Bonesteel~\cite{b89a} on a  square  lattice a $\pi/2$
rotation  exchanges sector $(+,-)$  and  sector $(-,+)$ but  sectors
$(-,-)$ and $(+,+)$ remain inequivalent.   A similar phenomenon occurs
on the triangular lattice  where $2\pi/3$ rotations permute cyclically
3 of the  4 sectors.  Consider a  finite-size triangular  lattice with
periodic boundary  conditions which has the $2\pi/3$-rotation symmetry
($N=L\times L$ for instance).  One draws 3 cuts $\Delta_{1,2,3}$ which
can be  deduced from each other  by $\pm2\pi/3$ rotations.   Of course
only two of  these three cuts are topologically  distinct but one  can
associate 3 parities  to  any short-rang dimer  configurations anyway.
These three cuts separate the sample in two sets of  $N/2$ sites.  If,
for instance, $N/2$ is even  the only allowed parities are  $(+,+,+)$,
$(+,-,-)$  $(-,+,-)$  and $(-,-,+)$  and  we recover  4  sectors as it
should be.  In this form it is now obvious that the three last sectors
are  symmetric   by   $2\pi/3$  rotations  and  will    have the  same
ground-state  energy.  One  can  also easily  determine the  relations
between the quantum numbers of these triplets.  As result, if $N/2$ is
even, sectors $(+,-)$, $(-,+)$  and $(-,-)$ will be exactly degenerate
and  if $N/2$  is  odd $(+,-)$, $(-,+)$ and   $(+,+)$  will be exactly
degenerate

If we consider how    $2\pi/3$ rotations and  translations  act  on this
three-dimensional  manifold, one can  easily   show that  this  3-fold
degeneracy can only take two    forms.  a) The degenerate states   can
correspond     to  a three-dimensional     representation.   The  only
possibility  is     that    the  corresponding  momentum  are    ${\bf
K}_{i=1,2,3}$.   b) The  other  possibility is  that a two-dimensional
representation      is    degenerate       with  a     one-dimensional
representation. The    only  possible realization  is    a ${\bf k}=0$
$R_{2\pi/3}=1$ state degenerate with   two ${\bf k}=0$  $R_{2\pi/3}=j$
and  $j^2$ complex conjugate states.  For  these samples, the $2\pi/3$
rotation  plays exactly the same role  as  the one-step translation in
the case of even$\times$odd samples.

All these properties were checked numerically in the QHCD model on the
triangular lattice.  We found  that the second  case of  degeneracy is
realized  for  this  model  in  the $N=28$   sample  (i.e.  the   four
eigenstates of the  quasi-degenerate ground-state multiplet have  wave
vector  ${\bf k}= (0,0)$ but belongs  to different IRs  of $C_3$).  In
fact, it must  be so for  any even$\times$even sample  at the RK point
since the  wave functions clearly   have  ${\bf k}=0$  in each  sector
(equal amplitude superposition of all  dimer coverings). This remains
true as long as the gap does  not close and should  be verified in the
whole spin-liquid phase of  the model.  In  that case both Proposals B
and B' of subsection~\ref{1DLSMA} are incorrect.

We can  ask what is the  result of Oshikawa's  insertion of a  flux in
such an  even$\times$even  spin-liquid. When  boundary  conditions are
twisted, the Hamiltonian is no longer $R_{2\pi/3}$-symmetric and the 2
irreducible representations in which the topological ground-states lie
($R_{2\pi/3}=1$ and $R_{2\pi/3}=j,j^2$) merges  into a single one with
momentum  ${\bf k}=(0,0)$.  For this   reason,   during the  adiabatic
process, these four energy  levels  cannot cross and must  necessarily
map onto  themselves  after the complete flux   quantum insertion.  In
that  case Oshikawa's    procedure    do not   relate   the  different
ground-states with each other.

\end{subsection}

\begin{subsection}{4-fold degeneracy in RVBSL phases}
\label{sec:4fdeg}

The numerical   data  on MSE  model    suggest that  the  ground-state
degeneracy is  4  in  a  spin  liquid phase~\footnote{A  ground-state
degeneracy has been evoked  by Wen and co-workers~\cite{wwz89} in {\em
chiral}  spin liquids.  To  our   understanding, this is  a  bit  more
complex than the present case. A chiral spin  liquid is a gapped phase
with  long-range order in  chirality  (local observable defined as the
triple product of 3 neighboring spins).  It not only breaks reflection
symmetry but  also  time reversal  symmetry and it  is supposed  to be
$2(k)^d$ degenerate,  where $k$  is   an even  index  related to   the
fractional statistics of the elementary excitations.}, whatever may be
the   shape of the   sample  (Fig.~\ref{gaps_size}).  In  the case  of
even$\times$odd sample ($N=30=6\times  5$ in Table~\ref{table-3}), the
quasi-degenerate     levels   exactly   satisfy    the   constraint  of
Eq.~(\ref{symmetryqn}).   In  even$\times$even samples which  have the
$2\pi/3$-rotation symmetry ($N=28$  and $36$ in  Table~\ref{table-3} and Fig.~\ref{sp36}),
the quasi-degeneracy is observed  between a ${\bf k}=(0,0)$  state and
three  ${\bf k}={\bf K}_{1,2,3}$,    in  perfect agreement   with  the
arguments  developed in paragraph~\ref{sss:exact_deg_e_e}.  These  are
additional arguments in favor of the topological interpretation of the
quasi-degeneracy observed in this model\cite{mlbw99}.

\begin{figure}[b]
        \begin{center}
        \resizebox{8cm}{!}{
        \includegraphics{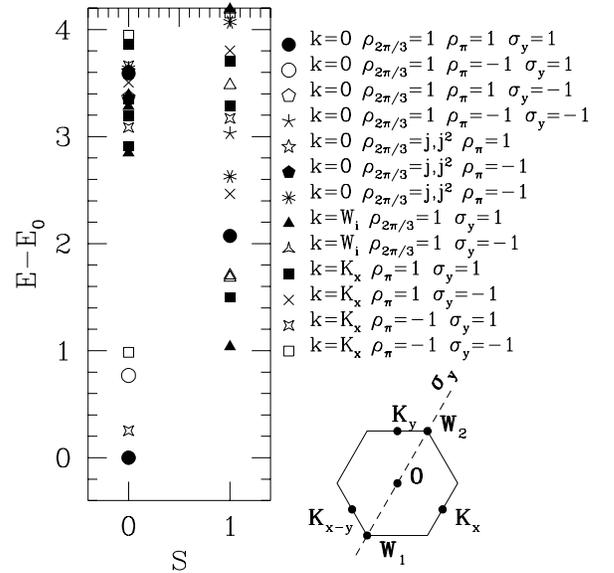}}
        \end{center}

        \caption[99]  {First eigenstates of the multiple-spin exchange
        model on a $6 \times  6$ sites sample (parameters $J_2=-2$ and
        $J_4=1$; the system is in a spin liquid phase). The first four
        levels  with  total spin $S=0$  and  some $S=1$ are displayed.
        The quantum numbers  of the eigenstates  are displayed  at the
        right of the figure.   The ground-state belongs to the trivial
        representation  of the group   of translations, rotations  and
        reflections  (in particular  ${\bf k}=(0,0)$, $\rho^\pi_0  =1$
        and   $\sigma^y_0=1$).    The  first $S=0$ excited    state is
        three-fold    degenerate     (wave-vectors      ${\bf   K}_i$,
        $\rho^\pi_1=-1$ and $\sigma^y_1=1$).  The finite-size  scaling
        indicates   that    this  state collapses   to    the absolute
        ground-state  in the thermodynamic limit  (Table~\ref{table-2}
        and Fig.~\ref{gaps_size}). The  third  and fourth energies  in
        the  $S=0$ spin  sector are  also probably  degenerate in  the
        thermodynamic limit (global 4-fold degeneracy).}
\label{sp36}
\end{figure}

\begin{table*}[t]
        \begin{center}\begin{tabular}{|c|c|c|c|c|}
        \hline
        N&28&30&32&36\\
        \hline
${\bf k_0}=(0,0)$&-110.965 (1)&-117.896 (-1)&-126.608(1,1)&-142.867 (1,1)\\
${\bf k_1}=(\pi,0)$&-109.396 (-1)&-117.193 (1)&-125.684(-1,-1)&-142.615 (-1,1)\\  
        \hline
${\bf k_0}=(0,0)$&-108.168 (1)&-116.590 (1)&*&-142.098 (-1,1)\\
${\bf k_1}=(\pi,0)$&-107.633 (-1)&-116.544 (-1)&*&-141.883 (-1,-1)\\
        \hline
        \end{tabular}\end{center} 

        \caption[99]{Energies  and  quantum numbers  of  the first two
        multiplets  in the spin liquid  spectra of the  MSE model. The
        wave vector  is indicated in   the first column,  the  quantum
        numbers $\rho^{\pi}$     and  $\sigma^y$  are     indicated in
        parenthesis after the energy.  Samples $N=28$ and $30$ have no
        axial symmetry.  Relations~(\ref{symmetryqn}) are satisfied in
        the even-odd .   * stands  for  data  where the 2$^{\rm   nd}$
        multiplet does not appear clearly, due to the absence of $C_3$
        symmetry of the sample.}
        \label{table-3}
\end{table*}

The  same situation appears in  the liquid phase of  the QHCD model on
the triangular  lattice (we consider the  $J=1$  and $V=0.95$ point
 to avoid difficulties with the first order transition to
the  staggered phase at $V=1$).  This  degeneracy property was already
inferred     by various      authors       at  the  end   of       the
eighties~\cite{s88,rk88,rc89}.       From            the   results  of
section~\ref{subsec:4sectors} we  know  that the  ground-states inside
each of the 4 topological sectors give  rise to two energy levels (two
doublets  or one  singlet   plus a  triplet  depending on  the  sample
symmetry).  Indeed a quasi four-fold degeneracy of the ground-state is
observed.  More  precisely, the two levels  approach each other as the
linear size  is increased.  For instance,  this  small level splitting
drops  from $\delta=0.02285$ on  a  $16=4\times4$ sample to $0.008147$
for   $30=6\times5$   sites.  In   samples   without $2\pi/3$-rotation
symmetry a  set    of  four  nearly  degenerate   levels  is  observed
($N=24=6\times 4$ for instance).

We make the following assumption: a) the ground-state can be described
in  a  short-ranged  dimer   basis.   b)  All $n$-dimer   correlations
($n=2,3,\cdots$)  are short-range  and  the corresponding  correlation
lengths are bounded.  c) The Hamiltonian  is local. From hypotheses a)
and c) it is clear that for a large enough system the four topological
sectors are  not  mixed in the ground-state  and  the spectrum  can be
computed separately in  each  sector.  We  do  not  have any  symmetry
operation which connects all {\em four} sectors and we need a physical
argument to explain the fact that energies should  be the same in each
sector (in  the thermodynamic  limit).   Because of their  topological
nature,  it is   not possible to  determine  to  which sector  a dimer
configuration belongs  by looking only at a  {\em finite area}  of the
system.  In other words, any dimer configuration  defined over a large
but finite   part of the    system  can be  equally  realized  in  all
sectors. The Hilbert  space available to the  system is the  same over
any finite  region  of the  system.   In the  absence of any  form  of
long-range  order  the    system  can  therefore   optimize   all  its
correlations  with  an arbitrary high   accuracy equally well  in each
sector.
At this point we can only conclude that  the four sector will have the
energy  {\em  density} and we  cannot  exclude the existence of  a gap
between  the  different  topological sectors.   However the  numerical
results obtained in the  QHCD and MSE models  indicate that it is  not
the case  and that the four  ground-state have asymptotically the same
{\em total} energy.  We  think that this should be  true for a general
short-range  RVB spin liquid.   Based on  the relationship between the
effect of a twist  and the topological  degeneracy in such  systems, a
vanishing gap between sectors in the  thermodynamic limit is likely to
be  related to the   complete absence of   sensitivity to such a twist
(section~\ref{GapRobustness}).

\end{subsection}

\subsection {Miscellaneous remarks on the RVB ground-state degeneracy}
\label{remarks}

--- {\em   Dimers     and twist  operator}.   The    variational state
$\left|\psi_{1,\Delta}\right>$   can   be    deduced  simply      from
$\left|\psi_0\right>$ by changing the  sign of the dimers crossing the
cut $\Delta$. Such an  operation can also been seen  as a $2\pi$ twist
of the spins of column 0. As was  noticed by Bonesteel~\cite{b89a} for
the  one-dimensional problem   such  an  operation  is  mathematically
related to action of the LSMA twist operator.  From the physical point
of   view   the   reason   why    $\left|\psi_{1,\Delta}\right>$   has
asymptotically the same energy as $\left|\psi_0\right>$ becomes clear:
in the absence of stiffness and  of long-range spin-spin correlations,
the  perturbation  induced by  the boundary  term  of Eq.~(\ref{Hphi})
cannot propagate and does not change  the energy of the initial state:
its only  effect is  to change  the  relative phases of the  different
topological  components of the   wave function,  and consequently  the
momentum  and space symmetry quantum  numbers of  the initial state of
the even-odd samples.

---  {\em Fractionalization    and topological  degeneracy.}  To   our
knowledge  all  present theoretical  descriptions  of   fractionalized
excitations       in        2D        magnets      or          related
problems~\cite{wen91,s01,sp01,k01,msf01})  (we should   also   mention
topological  properties of  Laughlin's   wave function for  fractional
quantum Hall effect~\cite{haldane85,wn90})     imply       topological
ground-state degeneracies.  In  such pictures, the physical  operation
which transforms a ground-state into another is the virtual creation of
a pair  of spinons (by  dimer breaking)  followed by  its annihilation
after the  circulation of  one of them   around the torus.   In such a
process a $\pi$  phase-shift   is introduced between   the topological
sectors (as in  the above recipe).  For samples  with an odd number of
rows  this operation connects eigenstates    with different ${\bf  k}$
vectors    (and    space   quantum     numbers)   as    described   in
Eqs.~(\ref{symmetryqn}).  Oshikawa's  adiabatic construction is of the
same nature.

--- {\em Numerical studies with a  different topology.} An interesting
check  of the  pure  topological nature of  this   degeneracy could be
obtained by studying the problem  no more on  a torus but on a surface
with  a  different genus.   On  a  sphere  we   expect an  absence  of
degeneracy.   Unfortunately if a  lattice   can be represented on   an
infinite plane, both the number of links $L$ and plaquettes $P$ depend
linearly on the  number of sites  $N$ and Euler's relation $P-L+N=2-G$
constrains the genius  $G$ to be 2~!  The  torus is the  only possible
topology   if we  require   a  full  translation invariance   in  both
directions.   In a  recent  work, Ioffe {\it  et al.}~\cite{if01} have
studied the absence of sensitivity    to disorder as an evidence   for
topological phenomenon in  the liquid phase of  the QHCD model  on the
triangular lattice.  They also used  {\em open} boundary conditions to
modify  the topology of  the system  and argue  in  that case that the
low-energy spectrum is free of edge states which could hide the actual
ground-state degeneracy.

--- {\em Example of RVB  phase with 2 spins  in the unit cell.} A spin
liquid  state, seeming very similar  to the state  observed in the MSE
model  on the triangular lattice, has  been observed  in the $J_1-J_2$
model on the hexagonal lattice~\cite{fsl01} for $J_1=-1,\;J_2=0.3$. No
quasi-degeneracy of the ground-state  has been  noticed. It should  be
remarked that in  this  system  there are  2 spins   $\frac{1}{2}$ per
crystallographic unit cell and no  degeneracy is expected on the basis
of the topological arguments.

\begin{subsection}{Symmetry breaking in gapped phases}
\label{sec:no_sym_break}

From the mathematical point of  view, ground-state wave functions that
break one-step translations or  space  group symmetries can  be  built
from linear  combinations of the degenerate  ground-states of the {\it
even-odd} samples.  In a completely  equivalent way, ground-state wave
functions    that  break rotation     symmetry    can  be   built   in
even$\times$even samples.  One could thus  superficially conclude that
spontaneous symmetry breaking is possible in RVBSL, we will show below
that this assumption is false.

There are many  features which show that this degeneracy property is a
subtle one, both from the mathematical and physical viewpoints.

The  possible alternation of the spatial  properties  of the low-lying
excitations with the  parity of the number of  rows of the sample
(as observed in the QHCD model on the triangular lattice) is a
first   difficulty.  The degeneracy  of  the  RVBSL  is in fact  quite
different from that  appearing in  a VBC.   We do  not expect the  VBC
ground-state degeneracy  to depend on the  genus of the sample, as the
RVBSL does.

From the physical   point of view  also  the two  situations are quite
different. An infinitesimal  symmetry breaking perturbation is able to
select one  symmetry breaking ground-state of the  VBC, but as we will
show below this is impossible in the RVBSL.

Let us call ${\cal A}$ the extensive non-diagonal observable
appearing in the VBC in the thermodynamic limit. On a columnar
VBC modulated in the {\bf u} direction, this observable is: 
\begin{equation}
	{\cal A}=
	\sum_{j=1}^{N}  e^{{i{\bf K}_1} \cdot {\bf r}_j}
	P_{S=0}( {\bf r}_j, {\bf r}_j + {\bf u})   
\end{equation}
 where $ P_{S=0}( {\bf r}_j, {\bf r}_j + {\bf u})$ is the
projector on the singlet state of two neighboring spins.
${\cal A}$ connects eigenstates with wave-vector ${\bf k}_0$ 
to states with wave-vector
${\bf k}_0 + {\bf K}_1$ . 

On a  finite size  sample,   with periodic  boundary conditions,   the
expectation value of ${\cal A}$ is zero in any eigenstate,
but  $<{\cal A}^2>$   could be non zero.  If
the order parameter ${\cal P}$ defined by:
\begin{equation}
{\cal P}^2= <\psi_{g.s.}|{\cal A}^{\dagger}{\cal A}|\psi_{g.s.}> /N^2
\end{equation}
 does  not  vanish  in the thermodynamic
limit, the system has columnar dimer long-range order with wave vector
${{\bf K}_1}$.

Let us now consider a perturbation of the Hamiltonian:
\begin{equation}
H_{\delta} = H_0 - \left(\delta {\cal A} +
h. c.\right).
\end{equation}
At T=0, the intensive linear response on the observable ${\cal A}$ is
measured by the susceptibility:
\begin{equation}
\chi = \frac{2}{N} <\psi_{g.s.}|{\cal A}^{\dagger}
\frac {1 }{  H_0 - E_{g.s} }{\cal A} |\psi_{g.s.}>
\label{suscep}
\end{equation}

This  susceptibility is bounded from below~\cite{ssgpt99}:
\begin{equation}
\frac {4 {\cal P}^4 N^2}{f}< \chi 
\label{ineq}
\end{equation}
where $f$ is the oscillator strength:
\begin{equation}
f = \frac{1}{N} \left<\psi_{g.s.}| \left[{\cal A},\left[H_0 ,{\cal A}\right]\right]|\psi_{g.s.}\right>
\end{equation}
The demonstration uses the properties of the spectral
decomposition associated to the operator ${\cal A}$:
\begin{equation}
S(\omega) = \frac {1 }{N}  \sum_{n \neq 0}\left|<\psi_{g.s.}|{\cal A}|n> \right|^2 \delta (\omega -\omega_n)
\end{equation}
where $\omega_n =E_n -E_{g.s}$
\begin{equation}
{\cal P}^2= \frac {1 }{N} \int S(\omega) d\omega
\end{equation}
Using the Cauchy Schwartz inequality one obtains:
\begin{equation}
{\cal P}^4 \leq \frac {1 }{N^2} \int \omega S(\omega)d\omega \int \omega^{-1} S(\omega)d\omega
\end{equation}
where  
\begin{equation}
\int \omega S(\omega)d\omega = f/2
\end{equation}
\begin{equation}
\int \omega^{-1} S(\omega)d\omega = \chi /2
\end{equation}
which proves inequality~(\ref{ineq}).
For a short-range Hamiltonian the oscillator strength $f$ is
$ {\cal O}(1)$ and  inequality (\ref{ineq}) implies that  
the T=0 susceptibility associated to a finite order parameter
diverges at least as the square of the sample
size: any infinitesimal  symmetry
breaking perturbation will select a symmetry
breaking state.

We will now show that for a RVBSL, where all the correlations
functions are short-ranged with  correlation lengths bounded by $\xi$,
the susceptibilities of the medium remain finite in the
thermodynamic limit. To do so we distinguish in Eq.~\ref{suscep}
the contributions from the quasi-degenerate states of the
topological multiplet (called $|\alpha_i>$)
 from the contribution of the other states of the spectrum,
above the physical gap $\Delta$. We thus obtain the following upper
bound for the susceptibility:
\begin{eqnarray}
\chi\; &=& \chi_{\rm top}\; + \chi_{\Delta} \\
\chi_{\rm top}\; &=&\frac{2}{N} <\psi_{g.s.}|{\cal A}^{\dagger}\frac{ 
|\alpha_1><\alpha_1| }{E_{\alpha_1} - E_{g.s.}}{\cal A} |\psi_{g.s.}>
\label{chitop}
\\
\chi_{\Delta} &\leq& \frac{2}{N \Delta } \left[ <\psi_{g.s.}|{\cal A}^{\dagger}{\cal A} \right.
|\psi_{g.s.}> \nonumber \\
 &-&<\psi_{g.s.}|{\cal A}^{\dagger}|\alpha_1><\alpha_1|{\cal
A}|\psi_{g.s.}>]
\label{chidelta}
\end{eqnarray}
where   $|\alpha_1>$ stands for   the    state(s) of the   topological
multiplet connected to the absolute ground-state by ${\cal A}$.  Using
the local properties  of ${\cal A}$, ${\cal  A}^{\dagger}|\alpha_1>$ is
in the same topological sector as $|\alpha_1>$ and $<\psi_{g.s.}|{\cal
A}^{\dagger}  |\alpha_1>$ is at  most of ${\cal O}(N \times 2^{-L/2})$
(see  paragraph~\ref{deftopsec}).   As  $E_{\alpha_1}  - E_{g.s.}$  is
supposed to   decrease  as  $exp(-L/\xi)$  (see  Fig.~\ref{gaps_size})
$\chi_{\rm top}$  goes to a constant when the
size of  the   sample  goes to   infinity,  provided $\xi$   is  small
enough~\footnote{Strictly  speaking $\xi$ should   be $\leq\log 2$ but
going   from    Eq.~\ref{ineqA}    to  Eq.~\ref{ineqB}    dramatically
overestimates the  scalar product in the case  of an RVBSL. The reason
is    that   if  two   dimer     coverings  $c^+$   and $c^-$ maximize
$\left<c^+|c^-\right>$  they only differ along  a  single large ($\sim
L$) loop. They have different  local  correlations along the loop  and
their energy difference is of order $L$  and it is very unlikely that
their weights in the ground states $\psi^+(c^+)$ and $\psi^-(c^-)$ are
{\em both} of order one.}.
In a  system with exponentially  decreasing correlations, ${\cal P}^2$
decreases as $1/N$ and $\chi_{\Delta}$ is trivially constant at the
thermodynamic limit.

In such a phase  an  infinitesimal field  cannot    induce a symmetry
breaking and   there  could not  be any {\it spontaneous}
symmetry breaking.

\end{subsection}

\end{section}

\begin{section} {Conclusion}
\label{section:conclusion}

The main statements of this paper could be summarized in the following
way. In any situations the ground-state of an $SU(2)$ Hamiltonian with
one  spin-$\frac{1}{2}$ per   unit    cell   is degenerate    in   the
thermodynamic limit.  In   fact   the problem encompasses  two    very
different cases:
\begin{itemize}
	\item  Either the  system has  a $T=0$  macroscopic long-range
	order: in these  situations the degeneracy of the ground-state
	encompasses  and  reveals   the  symmetry   breakings  of  the
	macroscopic ground-states. It is  algebraic in the system size
	in case of a continuous $SU(2)$ symmetry breaking and involves
	only a finite number of levels in case  of a discrete symmetry
	breaking.    This    situation   is    not    restricted    to
	spin-$\frac{1}{2}$ Hamiltonians.

	\item Or the system has no long-range order  at $T=0$, it is a
	spin  liquid.   In that  case the  degeneracy property depends
	essentially on the existence of an odd-half-integer total spin
	in the unit cell. In this paper we  have only studied the case
	of  a unique spin-$\frac{1}{2}$ per unit  cell, but in view of
	its  topological origin we  suspect the result  to be true for
	the general case  above. On the other hand,  we have given one
	example with two spins $\frac{1}{2}$ in the unit cell and very
	plausibly   a    unique  ground-state  in   the  thermodynamic
	limit~\cite{fsl01}.
\end{itemize}

All these ideas had already been known by a number of authors.
The really new contributions of this work are the following:
\begin{itemize}
	\item  We  have   analyzed  the Lieb   Schultz  Mattis Affleck
	conjecture for  2D  spin systems and   shown  by the numerical
	study of a  counter-example that a part  of this conjecture is
	probably incorrect (section~\ref{section:2DLSMA}).

	\item For spin liquid   systems,  we have  analyzed   Oshikawa
	adiabatic construction and shown that  it is equivalent to the
	hypothesis   of an    absence  of stiffness   of   the  gapped
	systems.  From our numerical analysis,  we conjecture that the
	spin  liquids display   in the  thermodynamic   limit a  total
	absence   of   sensitivity    to  the   boundary    conditions
	(section~\ref{section:Oshikawa}).

	\item We have shown that on  a 2D torus with  an odd number of
	rows, the   ground-state  is degenerate  (which  was   already
	known),   we  have also  demonstrated  some exact relationship
	between the quantum  numbers of these multiplets and  extended
	the  demonstration  of  degeneracy   to some  even$\times$even
	systems    with rotation symmetry    (which is  new).  We have
	developed  the idea that these  spatial properties of the spin
	liquid spectra depend crucially on the parity of the number of
	rows in the sample and have developed both exact and numerical
	examples to support  this  statement. The  study of   the QHCD
	model in even$\times$even samples is an explicit example where
	proposals B and B' of LSMA and Oshikawa are not verified.
	
	\item  Endly we have shown why,  in spite of the degeneracy of
	the   ground-state,    usual   spin   liquids  cannot  exhibit
	spontaneous  symmetry breaking of any  kind contrarily to VBCs
	(or chiral spin  liquid)  (the ultimate reason of  our obscure
	and sometimes controversed   previous claim that  the  invoked
	symmetry breakings in spin liquids were ``nonphysical'').
\end{itemize}

{\bf  Acknowledgments:} We acknowledge  very fruitful discussions with
S. Sachdev and V. Pasquier.  One of  us (C.L.)  thanks the hospitality
of I.T.P. (Santa Barbara) and the organizers  of the Quantum Magnetism
program.  Special thanks are  due to I.  Affleck  and D.  Mattis whose
vivid interest has prompted this work.  This research was supported in
part by the National Science  Foundation under Grant No.  PHY94-07194,
and by the CNRS and the Institut de  D{\'e}veloppement des Recherches en
Informatique          Scientifique         under             contracts
010091-010076.  Diagonalization of the  QHCD  model  were done on  the
Compacq alpha server of the CEA under project 550.
\end{section}

\begin{section}{APPENDIX A: Symmetries of spectra with twisted boundary conditions}
        
\begin{subsection}{Arbitrary twist angle $\phi$}

With periodic boundary conditions, spectra  at momentum ${\bf k}$  and
$-{\bf k}$ are identical because  the Hamiltonian commutes with $\Rpi$
the lattice rotation of angle $\pi$ about the origin.  In the presence
of a non zero twist, this rotation is no longer a symmetry~:
\begin{equation}
        \Rpi \Hphi \Rpi^{-1} = \tilde{H}_{-\phi}
\end{equation}
If one  defines the {\em spin  flip} $F$ by $F S^z_{\bf x}  = - S^z_{\bf x} F$,
then $\Rpi F$
commutes         with      $\Hphi$    and     $E_{\phi}({\bf
k},S^z)=E_{\phi}(-{\bf k},-S^z)$.

   For  an arbitrary twist  the only
symmetries of $\tilde{H}_\phi$ are~:
\begin{itemize}
        \item $\Rpi F$ 
        \item spin  rotations about the $z$  axis~:
        $e^{i\theta S^z_{\rm tot}}$
        \item Translations
        \item   Other spatial symmetries  which do   not change the $x$
        coordinates. The  lattice may have, for instance,
        an axis of  symmetry parallel to the $x$ direction.
\end{itemize}
 All these symmetry operators commute and the Hilbert space of $\Hphi$
can thus be analyzed in terms of tensorial products of
one-dimensional IRs. 
\end{subsection}
\begin{subsection}{ $\pi$-twist }
 For  $\pi$ twist the  spectrum has one
additional symmetry.
  Since $U(2n\pi) \tilde{H}_\phi     U(2n\pi)^{-1}
=\tilde{H}_{\phi + 2n\pi}$  for integer $n$ , one  finds:
\begin{equation}
        \left[U(2\pi) \Rpi \right]
        \tilde{H}_\phi
        \left[U(2\pi) \Rpi \right]^{-1} = \tilde{H}_{2\pi-\phi}
\end{equation}
and therefore
\begin{equation}
        \left[U(2\pi) \Rpi \right]
        \tilde{H}_\pi
        \left[U(2\pi) \Rpi \right]^{-1} = \tilde{H}_{\pi}
\label{URHpi}
\end{equation}
This  new symmetry induces extra degeneracies  in  the spectra.  Using
Eqs.~(\ref{URHpi}) and (\ref{k1k2}), one obtains:
\begin{equation}
        E_{\phi=\pi}(S_{\rm tot}^z,{\bf k}_1)
        = E_{\phi=\pi}(S_{\rm tot}^z,{\bf k}_2)
\end{equation}
when
\begin{equation}
        {\bf k}_1+{\bf k}_2
        = ( \pi \left[L_y\;{\rm mod}\;2\right] + 2\pi\frac{S_{\rm tot}^z}{ L_x}  ,0)
\end{equation}
If $S_{\rm tot}^z=0$ this condition reduces to: 
\begin{equation}
        {\bf k}_1+{\bf k}_2
        = \left\{
        \begin{array}{cc}
         (\pi , 0) & \;\;L_y \;${\rm odd}$ \\
        ( 0 , 0) & \;\;L_y \;${\rm even}$ \\
        \end{array}
        \right.
\end{equation}
Of  course,   these degeneracies are explicitly   verified  in all our
numerical spectra.

Finally, we address  the issue of   the property in the $F\Rpi$
transform of the
eigenstates which are degenerate at   $\phi=\pi$
(the quantum number associated to this property is 
  $\eta= \pm 1$).
 For this purpose  we
compute the  commutation of $F\Rpi$ with  $U(2\pi)$~:
\begin{eqnarray}
        F\Rpi U(2\pi) & = & U(2\pi) F\Rpi  \exp
                \left(
                2i\pi S^z_{\rm tot} + 2i\pi S^z_0
                \right) \\
        F\Rpi U(2\pi) & = & U(2\pi) F\Rpi (-1)^{L_xL_y + L_y}
\end{eqnarray}
where $S^z_0=\sum_{p=0}^{L_y-1}  S^z_{n=0,p}$ involves  a sum over all
the spins  of the column  0.    Since $\Rpi$ trivially  commutes  with
$\eta$ we find that   each   doublet of degenerate  eigenstates   obey
$\eta_1=(-1)^{L_xL_y + L_y} \eta_2$.

\end{subsection}
\end{section}
\begin{section}{APPENDIX B: The multiple-spin exchange model
and  the type I spin  liquid  phase} The multiple  spin exchange (MSE)
Hamiltonian  is an effective Hamiltonian  which describes the exchange
of localized  fermions on    a  lattice. In  terms of    multiple-spin
permutation  operators, the $2-$, $3-$  and  $4-$spin exchanges can be
reduced to :
\begin{eqnarray}\label{MSE}
H=J^{eff}_2 \sum_{
        \begin{picture}(17,10)(-2,-2)
                \put (0,0) {\line (1,0) {12}}
                \put (0,0) {\circle*{5}}
                \put (12,0) {\circle*{5}}
        \end{picture}
} P_{2}
+J_4 \sum_{
\begin{picture}(26,15)(-2,-2)
        \put (0,0) {\line (1,0) {12}}
        \put (6,10) {\line (1,0) {12}}
        \put (0,0) {\line (3,5) {6}}
        \put (12,0) {\line (3,5) {6}}
        \put (6,10) {\circle*{5}}
        \put (18,10) {\circle*{5}}
        \put (0,0) {\circle*{5}}
        \put (12,0) {\circle*{5}}
        \end{picture}
} \left( P_{4}+P_{4}^{-1}\right) 
\end{eqnarray}
where  $J^{eff}_2$ (resp.   $J_4$)    measures a  combination of   the
tunneling  amplitude of $2-$  and  $3-$spin exchanges (resp.  $4-$spin
exchange).  The permutation operators  can  be rewritten in  terms  of
usual  spin operators: $ P_{ij}  =  2 {\bf S}_{i}\cdot{\bf S}_{j} +1/2
$.  $   P_{1234}    +h.c.$ is    a   polynomial   of  degree   two  in
$\bf{S}_{i}\cdot\bf{S}_{j}$.   A complete expression  in terms of spin
operators can be found in Ref.~\cite{mlbw99}.   This Hamiltonian has a
spin-gap phase  in a wide  range  of ferromagnetic  $ J^{eff}_2$,  and
antiferromagnetic  $J_4$ couplings~\cite{lmsl00}.   The  spin   liquid
spectra  studied in  the    present work  are at   the  specific point
$J^{eff}_2=-2$ and $J_4=1$, where the spin gap is of order 1.

\end{section}


\end{document}